\documentclass[pre,twocolumn,amsmath,showpacs,superscriptaddress,aps,longbibliography]{revtex4-1}

\usepackage{amssymb}
\usepackage{graphicx}
\usepackage{grffile}
\usepackage[english]{babel}
\usepackage[caption=false]{subfig}

\usepackage{newfloat}
\DeclareFloatingEnvironment[
	fileext=loa,
	listname=List of Algorithms,
	name=ALG.,
	placement=t,
]{algorithm}
\usepackage{algpseudocode}
\algrenewcomment[1]{\hangindent=0.5\linewidth \hfill \parbox[t]{0.5\linewidth}{\raggedright \settowidth{\hangindent}{\(\triangleright\) }\(\triangleright\) #1}}

\usepackage{fixmath}
\usepackage{mathtools}

\usepackage{hyperref}
\hypersetup{
		linktocpage,
    colorlinks,
    citecolor=black,
    filecolor=black,
    linkcolor=black,
    urlcolor=black,
}

\usepackage[capitalise,poorman]{cleveref}
\newcommand{\sref}[1]{\protect\subref{#1}}
\newcommand{\crefrangeconjunction}{--}

\crefrangelabelformat{figure}%
       {#3#1#4\crefrangeconjunction#5\crefstripprefix{#1}{#2}#6}
\crefmultiformat{figure}%
  {\edef\crefstripprefixinfo{#1}Figs.~#2#1#3}%
  { and~#2\crefstripprefix{\crefstripprefixinfo}{#1}#3}%
  {, #2\crefstripprefix{\crefstripprefixinfo}{#1}#3}%
  {, and~#2\crefstripprefix{\crefstripprefixinfo}{#1}#3}
\Crefmultiformat{figure}%
  {\edef\crefstripprefixinfo{#1}Figs.~#2#1#3}%
  { and~#2\crefstripprefix{\crefstripprefixinfo}{#1}#3}%
  {, #2\crefstripprefix{\crefstripprefixinfo}{#1}#3}%
  {, and~#2\crefstripprefix{\crefstripprefixinfo}{#1}#3}
\crefrangemultiformat{figure}%
{Figs.~#3#1#4\crefrangeconjunction#5\crefstripprefix{#1}{#2}#6}%
{ and~#3#1#4\crefrangeconjunction#5\crefstripprefix{#1}{#2}#6}%
{, #3#1#4\crefrangeconjunction#5\crefstripprefix{#1}{#2}#6}%
{, and~#3#1#4\crefrangeconjunction#5\crefstripprefix{#1}{#2}#6}
\Crefrangemultiformat{figure}%
{Figs.~#3#1#4\crefrangeconjunction#5\crefstripprefix{#1}{#2}#6}%
{ and~#3#1#4\crefrangeconjunction#5\crefstripprefix{#1}{#2}#6}%
{, #3#1#4\crefrangeconjunction#5\crefstripprefix{#1}{#2}#6}%
{, and~#3#1#4\crefrangeconjunction#5\crefstripprefix{#1}{#2}#6}

\crefrangelabelformat{section}%
       {#3#1#4\crefrangeconjunction#5\crefstripprefix{#1}{#2}#6}
       
\crefmultiformat{section}%
  {\edef\crefstripprefixinfo{#1}Sections~#2#1#3}%
  { and~#2\crefstripprefix{\crefstripprefixinfo}{#1}#3}%
  {, #2\crefstripprefix{\crefstripprefixinfo}{#1}#3}%
  {, and~#2\crefstripprefix{\crefstripprefixinfo}{#1}#3}
\Crefmultiformat{section}%
  {\edef\crefstripprefixinfo{#1}Sections~#2#1#3}%
  { and~#2\crefstripprefix{\crefstripprefixinfo}{#1}#3}%
  {, #2\crefstripprefix{\crefstripprefixinfo}{#1}#3}%
  {, and~#2\crefstripprefix{\crefstripprefixinfo}{#1}#3}
\crefrangemultiformat{section}%
{Sections~#3#1#4\crefrangeconjunction#5\crefstripprefix{#1}{#2}#6}%
{ and~#3#1#4\crefrangeconjunction#5\crefstripprefix{#1}{#2}#6}%
{, #3#1#4\crefrangeconjunction#5\crefstripprefix{#1}{#2}#6}%
{, and~#3#1#4\crefrangeconjunction#5\crefstripprefix{#1}{#2}#6}
\Crefrangemultiformat{section}%
{Sections~#3#1#4\crefrangeconjunction#5\crefstripprefix{#1}{#2}#6}%
{ and~#3#1#4\crefrangeconjunction#5\crefstripprefix{#1}{#2}#6}%
{, #3#1#4\crefrangeconjunction#5\crefstripprefix{#1}{#2}#6}%
{, and~#3#1#4\crefrangeconjunction#5\crefstripprefix{#1}{#2}#6}

\newcommand{\notation}[1]{{#1}}
\newcommand{\define}[1]{\emph{#1}}
\newcommand{\quoting}[1]{``#1''}
\newcommand{\todefine}[1]{``#1''}
\newcommand{\paraphrase}[1]{`#1'}
\newcommand{\mesostructure}{meso-set}

\renewcommand{\vec}[1]{\mathbold{#1}}
\newcommand{\balpha}{\vec{\alpha}}
\newcommand{\bbeta}{\vec{\beta}}
\newcommand{\N}{\mathbb{N}}
\newcommand{\R}{\mathbb{R}}
\newcommand{\PP}{\mathbb{P}}

\newcommand{\LP}{\widetilde{P}}

\newcommand{\set}[1]{\mathcal{#1}}
\newcommand{\mat}[1]{\mathbold{#1}}
\newcommand{\comp}[1]{\overline{#1}}

\newcommand{\mean}[1]{\langle#1\rangle}
\DeclareMathOperator{\gamdist}{\Gamma}
\DeclareMathOperator{\NMI}{NMI}

\begin{document}

\title{A framework for the construction of generative models \\ for mesoscale structure in multilayer networks}

\author{Marya Bazzi}
\thanks{Both authors contributed equally.}
\affiliation{Oxford Centre for Industrial and Applied Mathematics, Mathematical Institute, University of Oxford, Oxford OX2 6GG, United Kingdom}
\affiliation{The Alan Turing Institute, London NW1 2DB, United Kingdom}
\affiliation{Warwick Mathematics Institute, University of Warwick, Coventry CV4 7AL, United Kingdom}

\author{Lucas G. S. Jeub}
\thanks{Both authors contributed equally.}
\affiliation{Oxford Centre for Industrial and Applied Mathematics, Mathematical Institute, University of Oxford, Oxford OX2 6GG, United Kingdom}
\affiliation{Center for Complex Networks and Systems Research, School of Informatics and Computing, Indiana University, Bloomington, Indiana 47408, USA}
\affiliation{ISI Foundation, Turin, Italy}

\author{Alex Arenas}
\affiliation{Departament d'Enginyeria Inform\`atica i Matem\`atiques, Universitat Rovira i Virgili, 43007 Tarragona, Spain}

\author{Sam D. Howison}
\affiliation{Oxford Centre for Industrial and Applied Mathematics, Mathematical Institute, University of Oxford, Oxford OX2 6GG, United Kingdom}

\author{Mason A. Porter}
\affiliation{Oxford Centre for Industrial and Applied Mathematics, Mathematical Institute, University of Oxford, Oxford OX2 6GG, United Kingdom}
\affiliation{{CABDyN} Complexity Centre, University of Oxford, Oxford OX1 1HP, United Kingdom}
\affiliation{Department of Mathematics, University of California, Los Angeles, Los Angeles, California 90095, USA}

%%%%%

\begin{abstract}

Multilayer networks allow one to represent diverse and coupled connectivity patterns --- e.g., time-dependence, multiple subsystems, or both --- that arise in many applications and which are difficult or awkward to incorporate into standard network representations. In the study of multilayer networks, it is important to investigate mesoscale (i.e., intermediate-scale) structures, such as dense sets of nodes known as communities, to discover network features that are not apparent at the microscale or the macroscale. The ill-defined nature of mesoscale structure and its ubiquity in empirical networks make it crucial to develop generative models that can produce the features that one encounters in empirical networks. Key purposes of such generative models include generating synthetic networks with empirical properties of interest, benchmarking mesoscale-detection methods and algorithms, and inferring structure in empirical multilayer networks. In this paper, we introduce a framework for the construction of generative models for mesoscale structures in multilayer networks. Our framework provides a standardized set of generative models, together with an associated set of principles from which they are derived, for studies of mesoscale structures in multilayer networks. It unifies and generalizes many existing models for mesoscale structures in fully-ordered (e.g., temporal) and unordered (e.g., multiplex) multilayer networks. One can also use it to construct generative models for mesoscale structures in partially-ordered multilayer networks (e.g., networks that are both temporal and multiplex). Our framework has the ability to produce many features of empirical multilayer networks, and it explicitly incorporates a user-specified dependency structure between layers.  We discuss the parameters and properties of our framework, and we illustrate examples of its use with benchmark models for community-detection methods and algorithms in multilayer networks.
\end{abstract}

\maketitle

%%%%%%%%%%%%%%%%%%%%%%%%%%%%%%%%
%%%%%%%%%%%%%%%%%%%%%%%%%%%%%%%% 

\section{Introduction}
\label{sec:introduction}

One can model many physical, technological, biological, financial, and social systems as networks. The simplest type of network is a graph \cite{Newman2018}, which consists of a set of nodes (which represent entities) and a set of edges between pairs of nodes that encode interactions between those entities. One can consider either unweighted graphs or weighted graphs, in which each edge has a weight that quantifies the strength of the associated interaction. Edges can also incorporate directions to represent asymmetric interactions or signs to differentiate between positive and negative interactions.

However, this relatively simple structure cannot capture many of the possible intricacies of connectivity patterns between entities. For example, in \define{temporal networks}~\cite{Holme2012, Holme2015}, nodes and/or edges change in time; and in \define{multiplex networks}~\cite{Kivela2014}, multiple types of interactions can occur between the same pairs of nodes. To better account for the complexity, diversity, and dependencies in real-world interactions, one can represent such connectivity patterns using \todefine{multilayer networks} (see \cref{sec:multilayer_networks})~\cite{Kivela2014, Boccaletti2014, Aleta2018, Bianconi2018,Porter2018}. We use the term \define{single-layer network} (which is also called a ``monolayer network'' or a ``graph'') for a multilayer network with a single layer (i.e., an \paraphrase{ordinary} network), and we use the term ``network'' to refer to both single-layer and multilayer networks. 

By using a single multilayer network instead of several independent single-layer networks, one can account for the fact that connectivity patterns in different layers often \todefine{depend} on each other. (We use the term \define{dependent} in a probabilistic sense throughout the paper: an object $A$ depends on $B$ if and only if $\PP(A | B) \neq \PP(A )$.) For example, the connectivity patterns in somebody's Facebook friendship network today may depend both on the connectivity patterns in that person's Facebook friendship network last year (temporal) and on the connectivity patterns in that person's Twitter followership network today (multiplex). Data sets that have multilayer structures are increasingly available (e.g., see Table 2 of~\cite{Kivela2014}). A natural type of multilayer network consists of a sequence of dependent single-layer networks, where layers may correspond to different temporal snapshots, different types of related interactions that occur during a given time interval, and so on. Following existing terminology~\cite{DeDomenico2014a, Porter2018}, we refer to an instance of a node in one \todefine{layer} as a \todefine{state node} (see \cref{sec:multilayer_networks}). 

There is a diverse set of models for multilayer networks. (We overview them in \cref{sec:existing_models}.) Many of these models take a specific type of dependency (e.g., temporal) as their starting point. In the present paper, we introduce a framework for the construction of generative models for multilayer networks that incorporate a wide variety of structures and dependencies. It is broad enough to unify 
many existing, more restrictive interlayer specifications, but it is also easy to customize to yield multilayer network models for many specific cases of interest. Key purposes of such generative models include (1) generating synthetic networks with empirical features of interest, (2) benchmarking methods and algorithms for detecting mesoscale structures, and (3) inferring structure in empirical multilayer networks.

%%%

\subsection*{A unifying framework}

A key feature of multilayer networks is their flexibility, which allows one to incorporate many different types of data as part of a single structure. In this spirit, our goal is to provide a general, unifying framework that enables users to construct generative models of multilayer networks with a large variety of features of interest in empirical multilayer networks by appropriately constraining the parameter space. We accomplish this in two consecutive steps. First, we partition the set of state nodes of a multilayer network. Second, we allocate edges, given a multilayer partition. 
We focus on modeling dependency at the level of partitions (as was done in~\cite{Ghasemian2015}), rather than with respect to edges. Additionally, we treat the process of generating a multilayer partition separately from that of generating edges for a given multilayer partition. This modular approach yields random structures that can capture a wide variety of interlayer-dependency structures, including temporal and/or multiplex networks, appearance and/or disappearance of entities, uniform or nonuniform dependencies between state nodes from different layers, and others. For a specified interlayer-dependency structure, one can then use any (single-layer or multilayer) network model with a planted partition (i.e., a \define{planted-partition network model}) to generate a wide variety of network features, including weighted edges, directed edges, and spatially-embedded layers. 

The flexibility of our framework to generate multilayer networks with a specified dependency structure between different layers makes it possible to (1) gain insight into whether, when, and how to build interlayer dependencies into methods for studying many different types of multilayer networks and (2) generate tunable \todefine{benchmark models} that allow a principled comparison for community-detection (and, more generally, \todefine{\mesostructure{}-detection}) tools for multilayer networks, including for complicated situations that arise in many applications (such as networks that are both temporal and multiplex) but thus far have seldom or never been studied. In many benchmark models, one plants a partition of a network into well-separated \todefine{\mesostructure{}s} (e.g., communities), and one thereby imposes a so-called \quoting{ground truth} (should one wish to use such a notion) \cite{Peel2016} that a properly deployed \mesostructure{}-detection method ought to be able to find. Benchmark networks with known structural properties can be important for (1) analyzing and comparing the performance of different \mesostructure{}-detection tools; (2) better understanding the inner workings of \mesostructure{}-detection tools; and (3) determining which tool(s) may be most appropriate in a given situation. 

One can also use our framework to generate synthetic networks with desired empirical properties, to generate null networks, and to explore \quoting{detectability limits} of mesoscale structures. (See, for example, the study of detectability thresholds in \cite{Ghasemian2015} for a model that is a special case of our framework.) With some further work, it is also possible to use our framework to develop models for statistical inference. In our concluding discussion (see \cref{sec:conclusion}), we suggest directions for how to apply our framework to the task of statistical inference.
Our intention in designing such a flexible framework is to ensure that the generative models that it provides remain useful as researchers consider progressively more general multilayer networks in the coming years.

%%%%

\subsection*{Paper outline}

Our paper proceeds as follows.  
In \cref{sec:background}, we give an introduction to multilayer networks, overview mesoscale structure in networks, and review related work on generative models for mesoscale structure. 
In \cref{sec:notation}, we formally introduce the notation that we use throughout the paper.
In \cref{sec:sampling_communities}, we explain 
how we generate a multilayer partition with a specified dependency structure between layers. We also give examples of how to constrain the parameter space of our framework to generate several different types of multilayer networks and discuss what we expect to be common use cases (including temporal structure~\cite{Sarzynska2014, Ghasemian2015, Pamfil2018} and multiplex structure~\cite{DeDomenico2014a, Pamfil2018}) in detail. 
In \cref{sec:sampling_networks}, we describe how we generate edges that are consistent with the planted partition. 
In \cref{sec:numerical_examples}, we illustrate the use of our framework as a way to construct benchmark models for multilayer community detection. 
In \cref{sec:conclusion}, we summarize our main results and briefly discuss possible extensions of our work to enable statistical inference on empirical multilayer networks with our framework. 

Along with this paper, we include code~\cite{bazzi2019} that users can modify to readily incorporate different types of \todefine{null distributions} (see \cref{nulldistribution}), \todefine{interlayer-dependency structures} (see \cref{generalcase}), and planted-partition network models (see \cref{sec:sampling_networks}). 
The model instantiations that one needs for generating the figures in \cref{sec:numerical_examples} are available at \url{https://dx.doi.org/10.5281/zenodo.3304059}.

%%%%%%%%%%%%%%%%%%%%%%%%%%%%%%%%
%%%%%%%%%%%%%%%%%%%%%%%%%%%%%%%%

\section{Background and related work}
\label{sec:background}

%%%%

\subsection{Multilayer networks}
\label{sec:multilayer_networks}

\begin{figure}
\includegraphics[width=\linewidth]{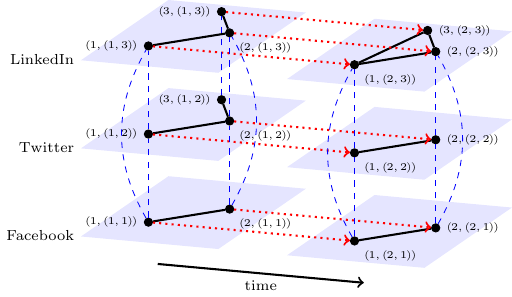}
\caption{Toy example of a multilayer network with three physical nodes and two aspects. We represent undirected intralayer edges using solid black lines, directed interlayer edges using dotted red lines with an arrowhead, and undirected interlayer edges using dashed blue arcs. The first aspect is ordered and corresponds to time. It has two time points (which we label as ``$1$'' and ``$2$''). Therefore, $\set{L}_1 = \{1,2\}$. The second aspect is unordered and represents a social-media platform. It has three elements: Facebook (labeled ``$1$''), Twitter (labeled ``$2$''), and LinkedIn (labeled ``$3$''). Therefore, $\set{L}_2 = \{1,2,3\}$. In this example, a state node takes the form $(i,(\alpha_1,\alpha_2))$, with physical node $i\in\{1,2,3\}$, aspect $\alpha_1\in\{1,2\}$, and aspect $\alpha_2\in\{1,2,3\}$. The total number of layers
 is $l = \vert \set{L}_1\vert \times \vert \set{L}_2\vert = 6$. 
}
\label{fig:notation}
\end{figure}

The simplest type of network is a graph $G = (\set{V},\set{E})$, where $\set{V} = \{1,\ldots,n\}$ is a set of nodes (which correspond to entities) and $\set{E} \subseteq \set{V}\times \set{V}$ is a set of edges. Using a graph, one can encode the presence or absence of connections (the edges) between entities (the nodes). However, in many situations, it is desirable to include more detailed information about connections between entities. A common extension is to allow each edge to have a weight, which one can use to represent the strength of a connection. We assign a weight to each edge using a \notation{weight function} $w: \set{E}\to \R$.

As we mentioned in \cref{sec:introduction}, one can also generalize the notion of a graph to encode different \define{aspects} of connections between entities, such as connections at different points in time or multiple types of relationships. We adopt the framework of multilayer networks~\cite{Kivela2014,DeDomenico2013, Bianconi2018, Aleta2018,Porter2018} to encode such connections in a network. In a \define{multilayer network}, a node can be present in a variety of different \define{states}, where each state is also characterized by a variety of different aspects. In this setting, edges connect
\define{state nodes}, each of which is the instantiation of a given node in a particular state. 
The set of all state nodes of a given entity corresponds to a single \define{physical node} (which represents the entity), and the set of all state nodes in a given state is in one \define{layer} of a multilayer network. In the remainder of this paper, we use the terms ``physical node'' and ``node'' interchangeably and the terms ``layer'' and ``state'' interchangeably. One can think of the aspects as features that one needs to specify to identify the state of a node. In other words, a state is a collection of exactly one element from each aspect.  For convenience, we introduce a mapping that assigns an integer label to each element of an aspect. That is, we map the $l_a$ elements of the $a^{\text{th}}$ aspect to the elements of a set $\{1,\dots,l_a\}$ of integers. Aspects can be unordered (e.g., social-media platform) or ordered (e.g., time). Most empirical investigations of multilayer networks focus on a single aspect (e.g., temporal~\cite{Bazzi2014, Betzel2017, Hric2018} or multiplex~\cite{cantini2015, Bacco2017, Stanley2015}). However, many real-world situations include more than one aspect (e.g., a multiplex network that changes over time). 
For an ordered aspect, we require that the mapping respects the order of the aspect (e.g., $t_i\to i$ for time, where $t_1\leq\ldots\leq t_{l_a}$ is a set of discrete time points). A multilayer network can include an arbitrary number of ordered aspects and an arbitrary number of unordered aspects, and one can generalize these ideas further (e.g., by introducing a time horizon) \cite{Kivela2014}.

To illustrate the above ideas, consider a hypothetical social network with connections on multiple social-media platforms (`Facebook', `Twitter', and `LinkedIn') between the same set of people at different points in time. In this example network, there are two aspects: social-media platform and time. 
(One can consider `type of connection' (e.g., `friendship' and `following') as a third aspect, but we restrict our example to two aspects for simplicity.)
The first aspect is ordered, and the number of its elements is equal to the number of time points or time intervals. (For simplicity, we often refer simply to ``time points'' in our discussions.) The second aspect is unordered and consists of three elements: `Facebook' (which we label with the integer ``$1$''), `Twitter' (which we label with ``$2$''), and `LinkedIn' (which we label with ``$3$''). If we assume that the time resolution is daily and spans the year $2010$, an example of a state is the tuple (`01-Jan-2010', `Twitter'), which is $(1,2)$ in our shorthand notation. We show a multilayer representation of this example social network in \cref{fig:notation}.

Edges in a multilayer network can occur either between nodes in the same state (i.e., \define{intralayer edges}) or between nodes in different states (i.e., \define{interlayer edges}). An example of an intralayer edge in \cref{fig:notation} is (1, (1, 1)) $\leftrightarrow$ (2, (1, 1)), indicating that entity $1$ 
is friends with entity $2$ on Facebook at time $1$. All interlayer edges in \cref{fig:notation} are \define{diagonal} (because they connect state nodes that correspond to the same physical node). Interlayer edges between layers at different times for a given social-media platform are \define{ordinal} (because they connect state nodes with successive time labels \footnote{It is also possible to have ordinal layering in aspects other than time.}) and directed. Interlayer edges between concurrent layers that correspond to different social-media platforms are \define{categorical} and undirected. (In this example, such edges occur between state nodes from all pairs of concurrent layers, although one can envision situations in which interlayer edges occur only between state nodes from a subset of such layers.)
An example of an ordinal intralayer edge in \cref{fig:notation} is (1, (1, 1)) $\rightarrow$ (1, (2, 1)), indicating a connection from entity $1$ in Facebook at time $1$ to entity $1$ in Facebook at time $2$. An example of a categorical intralayer edge in \cref{fig:notation} is (1, (1, 1)) $\leftrightarrow$ (1, (1, 2)), indicating a connection between node $1$ in Facebook at time $1$ and node $1$ in Twitter at time $1$. One can also imagine other types of interlayer edges in an example like the one in  \cref{fig:notation}, as there may be edges between state nodes at successive times and different social-media platforms (e.g., $(1,(1,1))\to(1,(2,2))$).

%%%%%%%%%%%%%%%%%%%%%%%%%%%%%%%%
%%%%%%%%%%%%%%%%%%%%%%%%%%%%%%%%

\subsection{Mesoscale structures in networks}
\label{sec:mesoscale_structure}

Given a (single-layer or multilayer) network representation of a system, it is often useful to apply a coarse-graining technique to investigate features that lie between those at the \define{microscale} (e.g., nodes, pairwise interactions between nodes, or local properties of nodes) and those at the \define{macroscale} (e.g., total edge weight, degree distribution, or mean clustering coefficient)~\cite{Newman2018}. One thereby studies \define{mesoscale} features such as community structure \cite{Porter2009,Fortunato2016,rosetti2018,cherifi2019}, core--periphery structure \cite{csermely2013, Rombach2017}, role structure \cite{rossi2015}, and others. 

Our framework can produce multilayer networks with any of the above mesoscale structures. Notwithstanding this flexibility, an important situation is multilayer networks with \todefine{community structure}, which are the most commonly studied type of mesoscale structure \cite{Newman2018}. Community detection is part of the standard toolkit for studying single-layer networks~\cite{Porter2009, Fortunato2010,Fortunato2016}, and efforts at community detection in multilayer networks have led to insights in applications such as brain and behavioral networks in neuroscience~\cite{muldoon2018}, financial correlation networks~\cite{Bazzi2014}, committee and voting networks in political science~\cite{Mucha2010b, Mucha2010}, networks of interactions between bacterial species~\cite{Stanley2015}, disease-spreading networks~\cite{Sarzynska2014}, social networks in Indian villages~\cite{omodei2016}, and much more. 

Loosely speaking, a \define{community} in a network is a set of nodes that are \quoting{more densely} connected to each other than they are to nodes in the rest of the network~\cite{Porter2009, Fortunato2010, Fortunato2016, Schaub2016, Jeub2014}. Typically, a \paraphrase{good community} should be a set of nodes that are \paraphrase{surprisingly well-connected} in some sense, but what one means by \paraphrase{surprising} and \paraphrase{well-connected} is often application-dependent and subjective. In many cases, a precise definition of ``community'' depends on the method that one uses to detect communities. In particular, many popular community-detection approaches in single-layer and multilayer networks define communities as sets in a partition of a network that optimizes a quality function such as modularity~\cite{Fortunato2016,Newman2018,newman2006}; stability~\cite{Petri2014, Lambiotte2009, Delvenne2010}; Infomap and its variants~\cite{DeDomenico2014a, Rosvall2008}; likelihood functions that are derived from stochastic block models (SBMs), which are models for partitioning a network into sets of nodes with statistically homogeneous connectivity patterns~\cite{Holland1983, Karrer2011, Stanley2015, Peixoto2015,kloumann2016}; and others~\cite{Kivela2014, Jeub2015, Wilson2016}. In this paper, we refer to a partition of the set of nodes of a single-layer network as a \define{single-layer partition} and to a partition of the set of state nodes of a multilayer network as a \define{multilayer partition}. The two primary differences between a community in a multilayer partition and a community in a single-layer partition are that (1) the former can include state nodes from different layers; and (2) \todefine{induced communities} (see \cref{sec:notation}) in one layer of a multilayer network may depend on connectivity patterns in other layers. For many notions of (single-layer or multilayer) community structure --- including the most prominent methods --- one cannot exactly solve a community-assignment problem in polynomial time (unless P = NP)~\cite{Fortunato2010, Brandes2008, Jacobs2014, Ghasemian2018}; and popular scalable heuristics currently have few or no theoretical guarantees on how closely an identified partition resembles an optimal partition. These issues apply more generally to the field of cluster analysis, such as in graph partitioning \cite{Schaeffer2007}, and many of the problems that plague community detection also apply to detecting other kinds of mesoscale structures.

Throughout our paper, to make it clear which results and observations apply to community structure in particular and which apply to mesoscale structure more generally, we use the term \quoting{community} when referring to a set in a partition of the set of nodes (or the set of state nodes) of a network that corresponds to community structure and the term \quoting{\mesostructure{}} (see \cref{sec:notation}) when referring to a set in a partition that corresponds to any type of mesoscale structure. (In particular, a community is a type of \mesostructure{}.)

%%%%%%%%%%%%%%%%%%%%%%%%%%%%%%%%
%%%%%%%%%%%%%%%%%%%%%%%%%%%%%%%%

\subsection{Generative models for mesoscale structure}
\label{sec:existing_models}

The ubiquity and diversity of mesoscale structures in empirical networks make it crucial to develop generative models of mesoscale structure that can produce features that one encounters in empirical networks. Broadly speaking, the goal of such generative models is to construct synthetic networks that resemble real-world networks when one appropriately constrains and/or calibrates the parameters of a model using information about a scenario or application. Generative models of mesoscale structure can serve a variety of purposes, such as (1) generating benchmark network models for comparing \mesostructure{}-detection methods and algorithms \cite{Lancichinetti2009, Lancichinetti2008, Granell2015, Ghasemian2015, Condon2000, Girvan2002}; (2) undertaking statistical inference on empirical networks \cite{Ghasemian2015, Karrer2011, Peixoto2018}; (3) generating synthetic networks with a desired set of properties~\cite{Jeub2018, Bacco2017}; (4) generating null models to take into account available information about an empirical network~\cite{Chen2016}; and (5) investigating \quoting{detectability limits} for mesoscale structure, as one can plant partitions that, under suitable conditions, cannot subsequently be detected algorithmically, despite the fact that they exist by construction \cite{decelle2011,Ghasemian2015, Taylor2016}. 

One of the main challenges in constructing a realistic generative model (even for single-layer networks) is the breadth of possible empirical features in networks. The available generative models for mesoscale structure in single-layer networks usually focus on replicating a few empirical features at a time (rather than all of them at once): heterogeneous degree distributions and community-size distributions~\cite{Lancichinetti2008, Karrer2011, Ormane2013}, edge-weight distribution~\cite{Lancichinetti2009, Aicher2014, Peixoto2018}, spatial embeddedness~\cite{Sarzynska2014, Newman2015}, and so on. Multilayer networks inherit all of the empirical features of single-layer networks, and they also have a key additional one: dependencies between layers. These interlayer dependencies can be ordered (as in most models of temporal networks), unordered (as in multiplex networks), combinations of these, or something more complicated.
However, despite this variety, existing generative models for mesoscale structure in multilayer networks allow only a restrictive set of interlayer dependencies. They assume either a temporal structure~\cite{Sarzynska2014,Granell2015,Ghasemian2015, Zhang2016, Hulovatyy2016, Pasta2016}, a simplified multiplex structure with the same planted partitions across all layers~\cite{Peixoto2015,Chen2015, Catala2016, Barbillon2016, Bacco2017}, or independent groups of layers such that layers in the same group have identical planted partitions~\cite{DeDomenico2014a, Stanley2015}. Using an alternative approach, a very recent model is able to generate multilayer partitions that satisfy the constraint that nonempty induced \mesostructure{}s in different layers 
that correspond to the same \mesostructure{} in the multilayer partition are identical \cite{Huang2019}. 
Recent work~\cite{Pamfil2018} on the link between multilayer modularity maximization and maximum-likelihood estimation of multilayer SBMs allows either temporal or multiplex 
interlayer dependencies with induced partitions that can vary across layers, but it makes restrictive assumptions on interlayer dependencies (e.g., all layers have the same set of nodes, interdependencies are \todefine{diagonal} and \todefine{layer-coupled}, and so on). Importantly, in all aforementioned generative models of mesoscale structure in multilayer networks, interlayer dependencies are either (1) not explicitly specifiable or (2) special cases of the framework that we discuss in the present paper. 

%%%%%%%%%%%%%%%%%%%%%%%%%%%%%%%%
%%%%%%%%%%%%%%%%%%%%%%%%%%%%%%%%

\section{Notation}
\label{sec:notation}

We now present a comprehensive set of notation for multilayer networks. Different subsets of notation are useful for different situations.

We consider a \notation{multilayer network} $M=(\set{V}_M,\set{E}_M,\set{V},\set{L})$ with $n = \vert \set{V}\vert$ \notation{nodes} {(i.e., \notation{physical nodes})} and $l=|\set{L}|$ \notation{layers} {(i.e., \notation{states}).} We use $d$ to denote the number of aspects and $\set{L}_a = \{1,\ldots,l_a\}$ to denote the labels of the $a^{\text{th}}$ \notation{aspect} (where $a \in \{1, \dots, d \}$). We use $\set{O}$ to denote the set of \notation{ordered aspects} and $\set{U}$ to denote the  set of \notation{unordered aspects}. For each ordered aspect $a\in\set{O}$, we assume that the labels $\set{L}_a$ reflect the order of the aspect. That is, for all $\alpha,\beta\in \set{L}_a$, we require that $\alpha < \beta$ if and only if $\alpha$ precedes $\beta$. We say that a multilayer network is \define{fully-ordered} if ${\set{U}=\emptyset}$, \define{unordered} if ${\set{O}=\emptyset}$, and \define{partially-ordered} otherwise. The set $\set{L}=\set{L}_1 \times \ldots \times \set{L}_d$ of states is the Cartesian product of the aspects, where a state $\balpha\in \set{L}$ is an integer-valued vector of length~$d$ and each entry specifies an element of the corresponding aspect. Note that $l = |\set{L}|= \prod_{a=1}^{d}l_a$.

We use $(i,\balpha)\in \set{V}_M\subseteq \set{V}\times \set{L}$ to denote the \notation{state node} (i.e., \quoting{node-layer tuple}~\cite{Kivela2014}) 
of physical node $i\in \set{V}$ in state $\balpha\in \set{L}$. We include a state node in $\set{V}_M$ 
if and only if the corresponding node exists in that state. The edges $\set{E}_M\subseteq \set{V}_M \times \set{V}_M$ in a multilayer network connect state nodes to each other. We use $((i,\balpha),(j,\bbeta))$ to denote a directed edge from $(i,\balpha)$ to $(j,\bbeta)$. For two state nodes, $(i,\balpha)$ and $(j,\bbeta)$, that are connected to each other via a directed edge $((i,\balpha),(j,\bbeta))\in \set{E}_M$, we say that $(i,\balpha)$ is an \define{in-neighbor} of $(j,\bbeta)$ and $(j,\bbeta)$ is an \define{out-neighbor} of $(i,\balpha)$. We categorize the edges into \notation{intralayer edges} $\set{E}_L$, which have the form $((i,\balpha),(j,\balpha))$ and link entities $i$ and $j$ in the same state $\balpha$, and \notation{interlayer} (i.e., \define{coupling}) \notation{edges} $\set{E}_C$, which have the form $((i,\balpha),(j,\bbeta))$ for $\balpha \neq \bbeta$. We thereby decompose the edge set as $\set{E}_M = \set{E}_L\cup \set{E}_C$. 
 
We define a weighted multilayer network by introducing a weight function $w: \set{E}_M \to \mathbb{R}$ (analogous to the weight function for weighted single-layer networks), which encodes the edge weights within and between layers. For an unweighted multilayer network, $w(e)=1$ for all $e\in \set{E}_M$. We encode the connectivity pattern of a multilayer network using an \define{adjacency tensor} $\mat{A}$, which is analogous to the adjacency matrix for single-layer networks, with entries
\begin{equation}\label{eq:adjacency_tensor}
	A_{i,\balpha}^{j,\bbeta}=
		\begin{cases} 
			w(((i,\balpha),(j,\bbeta)))\,, & ((i,\balpha),(j,\bbeta))\in \set{E}_M \,,\\
			0\,,& \text{otherwise\,.}
		\end{cases}
\end{equation}

Note that $G_M=(\set{V}_M,\set{E}_M)$ is a graph on the state nodes of the multilayer network~$M$. We refer to $G_M$ as the \define{flattened network} of $M$. The adjacency matrix of the flattened network is the \quoting{supra-adjacency matrix}~\cite{Kivela2014,DeDomenico2014b,DeDomenico2013,Gomez2013} of the multilayer network.  One obtains a supra-adjacency matrix by flattening \footnote{
Flattening (i.e., \paraphrase{matricizing})~\cite{Kivela2014} a tensor $\mat{T}$ entails writing its entries in matrix form. Consequently, starting from a tensor $\mat{T}$ with elements $T^{\bbeta}_{\balpha}$, one obtain a matrix $\protect\widetilde{\mat{T}}$ with entries $\protect\widetilde{T}_{\tilde{\alpha},\tilde{\beta}}$, where there are bijective mappings between the indices.} an adjacency tensor (see \cref{eq:adjacency_tensor}) of a multilayer network. The multilayer network and the corresponding flattened network encode the same information~\cite{Kolda2009,Kivela2014}, provided one keeps track of the correspondence between the nodes in the flattened network and the physical nodes and layers of the multilayer network. In a similar vein, aspects provide a convenient way to keep track of the correspondence between state nodes and, for example, temporal and multiplex relationships.

We denote a multilayer partition with $n_{\mathrm{set}}$ sets
by $\set{S} = \{\set{S}_1,\ldots,\set{S}_{n_{\mathrm{set}}}\}$, where $\bigcup_{s=1}^{n_{\mathrm{set}}}\set{S}_s = \set{V}_M$ and $\set{S}_s\cap \set{S}_r = \varnothing$ for $s\neq r$. 
We represent a partition $\set{S}$ using a \define{partition tensor} $\mat S$ with entries $S_{i,\balpha}$, where ${S_{i,\balpha} = s}$ if and only if the state node $(i,\balpha)$ is in the set $\set{S}_s$.  
We refer to a partition of a temporal network (a common type of multilayer network) as a \define{temporal partition} and a partition of a 
multiplex network (another common type of multilayer network) as a \define{multiplex partition}. A multilayer partition induces a partition $\set{S}\vert_{\balpha} = \{\set{S}_1\vert_\balpha,\,\ldots, \set{S}_{n_{\mathrm{set}}}\vert_\balpha\}$ on each layer, where $\set{S}_s\vert_\balpha = \{i\in \set{V}: (i,\balpha)\in \set{S}_s\}$. We refer to a set $\set{S}_s$ of a partition $\set{S}$ as a \define{\mesostructure{}}, to $\set{S}\vert_{\balpha}$ as the \define{induced partition} of $\set{S}$ on layer $\balpha$, and to $\set{S}_s\vert_\balpha$ as the \define{induced \mesostructure{}} of $\set{S}_s$ on layer $\balpha$.
A community is a set $\set{S}_s$ in a partition that corresponds to community structure. We call $s\in\{1,\ldots,n_{\mathrm{set}}\}$ the \define{label} of \mesostructure{} $\set{S}_s$. One way to examine overlapping \mesostructure{}s is by identifying multiple state nodes from a 
single layer with the same physical node. 

%%%%%%%%%%%%%%%%%%%%%%%%%%%%%%%%
%%%%%%%%%%%%%%%%%%%%%%%%%%%%%%%%

\section{Generating multilayer partitions}
\label{sec:sampling_communities}

The systematic analysis of dependencies between layers is a key motivation for analyzing a single multilayer network, instead of examining several single-layer networks independently. The goal of our partition-generation process is to model interlayer dependency in a way that can incorporate diverse types of dependencies. We now motivate our partition-modeling approach, and we describe it in detail in \cref{nulldistribution,generalcase,lessgeneral,specialcase}. 

The complexity of dependencies between layers can make it difficult to explicitly specify a joint probability distribution for \mesostructure{} assignments, especially for unordered or partially-ordered multilayer networks. To address this issue, we require only the specification of conditional probabilities for a state node's \mesostructure{} assignment, given the assignments of all other state nodes. The idea of conditional models is old and follows naturally from Markov-chain theory~\cite{Bartlett1978,Buuren2018}. Specifying conditional models (which capture different dependency features separately) rather than joint models (which try to capture many dependency features at once) is convenient for numerous situations. For example, conditionally specified distributions have been applied to areas such as spatial data modeling~\cite{Besag1974}, imputation of missing data~\cite{Buuren2018}, secure disclosure of information~\cite{Fienberg2005}, dependence networks for combining databases from different sources~\cite{Heckerman2000}, and Gibbs sampling~\cite{Gelfand2000, Hobert1998}. 

A problem with conditional models is that the different conditional probability distributions are not necessarily \define{compatible}, in the sense that there may not exist a joint distribution that realizes all conditional distributions~\cite{Hobert1998, Arnold1989, Chen2010, Chen2014, Heckerman2000}. Although several methods have been developed in recent years for checking the compatibility of discrete conditional distributions, these either make restrictive assumptions on the conditional distributions or have scalability constraints that significantly limit practical use~\cite{Kuo2011, Buuren2018}. Nevertheless, the employment of conditional models (even if potentially incompatible) is common~\cite{Kuo2017}, and they have been useful in many applications, provided that one cautiously handles any potential incompatibility~\cite{Buuren2018}. For our use case, potential incompatibility arises for unordered or partially-ordered multilayer networks. An issue that can result from it is the non-uniqueness of a joint distribution. We carefully design our partition-generation process such that it is well-defined irrespective of whether conditional distributions are compatible. In particular, we show that convergence is guaranteed and we address non-uniqueness by appropriately sampling initial conditions (see \cref{app:sampling}). We also suggest empirical checks to ensure that a generated partition reflects planted interlayer dependencies.

In our framework, we define conditional probabilities in two parts: we separately specify (1) independent layer-specific random components and (2) interlayer dependencies.  For a choice of interlayer dependencies, there are several features that one may want to allow in a multilayer partition. These include variation in the numbers and sizes of \mesostructure{}s across layers (e.g., \mesostructure{}s can gain state nodes, lose state nodes, appear, and disappear) and the possibility for these \mesostructure{} variations to incorporate features of the application at hand. (For example, in a temporal network, one may not want a \mesostructure{} to reappear after it disappears.) We ensure that such features are possible for any choice of interlayer dependency via independent layer-specific \define{null distributions} that specify the set of possible \mesostructure{} assignments for each layer and determine the expected size and expected number of \mesostructure{}s in the absence of interlayer dependencies.

Interlayer dependencies should reflect the type of multilayer network that one is investigating. For example, in a temporal network, dependencies tend to be stronger between adjacent layers and weaker for pairs of layers that are farther apart. It is common to assume that interlayer dependencies are uniform across state nodes for a given pair of layers (i.e., interlayer dependencies are ``layer-coupled''~\cite{Kivela2014}) and occur only between state nodes that correspond to the same physical node (i.e., interlayer dependencies are ``diagonal''~\cite{Kivela2014})~\cite{Mucha2010, Stanley2015, Taylor2016, Bassett2011, Bazzi2014, Sarzynska2014, Gomez2013, DeDomenico2014b}. However, these assumptions are too restrictive in many cases (e.g., situations in which dependencies are state-node-dependent or in which dependencies exist between state nodes that do not correspond to the same physical node)~\cite{DeDomenico2014a, Jeub2015, Buldu2017, Aslak2018, Valdano2015b, Barbillon2016}. To ensure that one can relax these assumptions, we allow 
dependencies to be specified at the level of state nodes (or at the level of layers, when a user assumes that dependencies are layer-coupled). We encode these interlayer dependencies in a user-specified \define{interlayer-dependency tensor} that determines the extent to which a state node's \mesostructure{} assignment in one layer \todefine{depends directly} on the assignment of state nodes in other layers (see \cref{generalcase,lessgeneral}). Our independence assumption on the null distributions allows us to encode all of the interlayer dependencies in a single object (an interlayer-dependency tensor). The entries of an interlayer-dependency tensor specify an \todefine{interlayer-dependency network} and correspond to the causal links for the flow of information (in the information-theoretic sense)
between different layers of a multilayer network. Therefore, they should reflect any constraints that one wishes to impose on the direct flow of information between layers. 
Longer paths in an interlayer dependency network yield
indirect dependencies between structures in different layers.

After specifying the interlayer-dependency structure, we define the conditional \mesostructure{} assignment probabilities such that we either sample a state node's \mesostructure{} assignment from the corresponding null distribution or obtain it by copying the assignment of another state node (based on the interlayer-dependency tensor). Using these conditional probabilities, we define an iterative update process on the \mesostructure{} assignments of state nodes to generate multilayer partitions with dependencies between induced partitions in different layers. When updating the \mesostructure{} assignments of state nodes, we respect the order of an aspect (e.g., temporal ordering). For a fully-ordered multilayer network, our update process reduces to sequentially sampling an induced partition for each layer based on the induced partitions of previous layers. 
For an unordered multilayer network, our update process defines a Markov chain on the space of multilayer partitions. We sample partitions from a stationary distribution of this Markov chain. This sampling strategy is known as \define{(pseudo\nobreakdash-)Gibbs sampling}~\cite{Gelfand2000,Hobert1998,Kuo2017, Heckerman2000}. (We use the word \quoting{pseudo} because the conditional probabilities that we use to define the Markov chain are not necessarily compatible.)
For a partially-ordered multilayer network, our update process combines these two sampling strategies.

In \cref{nulldistribution}, we describe possible choices for the independent layer-specific null distributions.
In \cref{generalcase}, we explain our framework for generating a multilayer partition with general interlayer dependencies. In \cref{lessgeneral}, we focus on the specific situation in which interlayer dependencies are layer-coupled and diagonal, and we also assume that a physical node is present in all layers (i.e., the network is ``fully interconnected'' \cite{Kivela2014}). In \cref{specialcase}, we illustrate the properties of example temporal and multiplex partitions that are generated by models that we construct from our framework. Additionally, we take advantage of the tractability of the special case of temporal networks to highlight some of its properties analytically.

%%%%%%%%%%%%%%%%%%%%%%%%%%%%%%%%
%%%%%%%%%%%%%%%%%%%%%%%%%%%%%%%%

\subsection{Null distribution}
\label{nulldistribution}

We denote the null distribution of layer $\balpha$ by $\PP_0^\balpha$ and the set of all null distributions by $\PP_0=\{\PP_0^{\balpha}, \balpha\in L\}$. A simple choice for the null distributions is a \define{categorical distribution}, where for each layer $\balpha$ and each \mesostructure{} label $s$, we fix the probability $p_s^{\balpha}$ that an arbitrary state node in layer $\balpha$ is assigned to a \mesostructure{} $s$ in the absence of interlayer dependencies. That is,
\begin{equation}
	\PP_0^\balpha[s]=\begin{cases} p_s^{\balpha}\, ,& s\in\{1, \ldots, n_{\mathrm{set}}\}\,,\\
	0\, , & \text{otherwise\,,}\end{cases} 
	\label{null}
\end{equation}
where $n_{\mathrm{set}}$ is the total number of \mesostructure{}s in the multilayer partition and $\sum_{s=1}^{n_{\mathrm{set}}} p_s^{\balpha}= 1$ for all $\balpha \in L$. 
The set $\{1, \ldots, n_{\mathrm{set}}\}$ is the set of \mesostructure{} labels, and the support of a null distribution is the set of labels that have nonzero probability.  In the absence of interlayer dependencies, a categorical null distribution corresponds to specifying independent multinomial distributions for the sizes of induced \mesostructure{}s on each layer and fixing the expected size $n p_s^{\balpha}$ of each induced \mesostructure{}. Therefore, by choosing the probabilities $p_s^{\balpha}$, one has some control over the expected number and sizes of \mesostructure{}s in a sampled multilayer partition.  A natural choice for $\vec{p}^{\balpha}$ is to sample it from a \define{Dirichlet distribution}, which is the conjugate prior for the categorical distribution \cite{Raiffa2000,Agarwal2010}. One can think of the Dirichlet distribution, which is the multivariate form of the beta distribution, as a probability distribution over the space of all possible categorical distributions with a given number of categories. Any other (probabilistic or deterministic) choice for $\vec{p}^{\balpha}$ is also possible. We give further detail about the Dirichlet distribution in \cref{app:nulldistribution}, 
where we discuss how one can vary its parameters to control the expected number and sizes of \mesostructure{}s in the absence of interlayer dependencies. This can allow a user to generate, for example, a null distribution with equally-sized \mesostructure{}s or a null distribution with a few large \mesostructure{}s and many small \mesostructure{}s. Furthermore, irrespective of the particular choice of categorical distribution, it may be desirable to have \mesostructure{}s that have a nonzero probability of obtaining state nodes from the null distribution in some, but not all, layers. In \cref{app:nulldistribution}, we give examples of how one can sample the support of the null distributions before the probability vectors to incorporate this property when modeling the birth and/or death of \mesostructure{}s in temporal networks and the {presence and/or absence} of \mesostructure{}s in multiplex networks.

In general, the choice of the null distributions can have a large effect on the set of sampled multilayer partitions and ought to be guided by one's use case. For our numerical examples in \cref{sec:numerical_examples}, we fix a value of $n_{\mathrm{set}} \in \{1, \ldots, nl\}$, where $n$ is the number of physical nodes in each layer and $l$ the number of layers (see \cref{sec:notation}); and we use a symmetric Dirichlet distribution with parameters $q=n_{\mathrm{set}}$ and $\theta=1$ (see \cref{app:nulldistribution}) to sample probability vectors $\vec{p}^{\balpha}$ of length $n_{\mathrm{set}}$. This produces multilayer partitions in which the expected \mesostructure{} labels are the same across layers (and given by $\{1,\ldots, n_{\mathrm{set}}\}$) in the absence of interlayer dependencies and for which the expected induced \mesostructure{} sizes (which are $np^\balpha_s$ in the absence of interlayer dependencies) 
differ across layers. 

%%%%%%%%%%%%%%%%%%%%%%%%%%%%%%%%
%%%%%%%%%%%%%%%%%%%%%%%%%%%%%%%%

\subsection{General interlayer dependencies}
\label{generalcase}
 
We denote the user-specified interlayer-dependency tensor by $\mat{P}$, where $P_{i,\balpha}^{j,\bbeta}$ is the probability that state node $(j,\bbeta)$ copies its \mesostructure{} assignment from state node $(i,\balpha)$, for any two state nodes ${(i,\balpha), (j,\bbeta)\in \set{V}_M}$, and where $\mat{P}$ is fixed throughout the copying process. The probability that state node $(j,\bbeta)$ copies its \mesostructure{} assignment from an arbitrary state node when state node $(j,\bbeta)$'s \mesostructure{} assignment is updated is
\begin{equation}
	\hat{p}_{j,\bbeta} = \sum_{(i,\balpha)\in \set{V}_M} P_{i,\balpha}^{j,\bbeta}\,,
\label{copying_prob} 
\end{equation}
where we require that ${\hat{p}_{j,\bbeta} \leq 1}$ for all state nodes ${(j,\bbeta)\in \set{V}_M}$. We also require that all intralayer probabilities are $0$; that is, ${P_{i,\balpha}^{j,\balpha}=0}$ for all ${i,j \in\set{V}}$ and ${\balpha \in \set{L}}$. We say that a state node $(j,\bbeta)$ \define{depends directly} on a state node $(i,\balpha)$ if and only if $P_{i,\balpha}^{j,\bbeta}$ is nonzero. By extension, we say that a layer $\bbeta$ depends directly on a layer $\balpha$ if there exists at least one state node in layer $\bbeta$ that depends directly on a state node in layer $\balpha$. The interlayer-dependency tensor induces the \define{interlayer-dependency network}, whose edges are all interlayer, directed, and pointing in the direction of information flow between layers. The in-neighbors (see \cref{sec:notation}) of a state node $(j,\bbeta)$ in this network consists of the state nodes from which $(j,\bbeta)$ can copy a \mesostructure{} assignment with nonzero probability. The support of the null distribution $\PP_0^{\bbeta}$ corresponds to the only possible \mesostructure{} assignments for a state node $(j,\bbeta)$ whenever the state node does not copy its assignment from one of its in-neighbors in the interlayer-dependency network.

For a given null distribution $\PP_0$ and interlayer-dependency tensor $\mat{P}$, a multilayer partition that results from our sampling process depends on four choices: (1) the way in which we update a state node assignment at a given step; (2) the order in which we update state node assignments; (3) the initial multilayer partition; and (4) the criterion for convergence of the iterative update process. We discuss points (1), (2), and (3) in the remainder of this subsection. We describe the sampling process in more detail and discuss convergence in \cref{app:sampling}. 

\subsubsection*{Update equation}

A single \mesostructure{}-assignment update depends only on the choice of state node to update and on the current multilayer partition. Let $\tau$ be an arbitrary update step of the copying process. Suppose that we are updating the \mesostructure{} assignment of state node $(j,\bbeta)$ at step $\tau$ and that the current multilayer partition is $\set{S}(\tau)$ (with partition tensor $\mat{S}(\tau)$). We update the \mesostructure{} assignment of state node $(j,\bbeta)$ either by copying the \mesostructure{} assignment in $\set{S}(\tau)$ from one of its in-neighbors in the interlayer-dependency network or by obtaining a new, random \mesostructure{} assignment from the null distribution $\PP_0^{\bbeta}$ for layer $\bbeta$. In particular, with probability $\hat{p}_{j,\bbeta} $, a state node $(j,\bbeta)$ copies its \mesostructure{} assignment from one of its in-neighbors in the interlayer-dependency network; and with probability $1 - \hat{p}_{j,\bbeta}$, it obtains its \mesostructure{} assignment from the null distribution $\PP_0^{\bbeta}$.  This yields the following update equation at step $\tau$ of our copying process: 
\begin{equation}\label{eq:updating}
\begin{aligned}
	&\PP[S_{j,\bbeta}(\tau+1) = s|\mat S(\tau)]\\[5pt]
	&\quad \qquad =\smashoperator{\sum_{{(i,\balpha)\in \set{V}_M}}}P_{i,\balpha}^{j,\bbeta}\,\delta(S_{i,\balpha}(\tau),s)\\[5pt]
	&\quad\qquad \qquad + \left({1-\hat{p}_{j,\bbeta}}\right)\PP_0^{\bbeta}[S_{j,\bbeta} = s]\,.
\end{aligned} 
\end{equation}

The update equation \eqref{eq:updating} is at the heart of our framework. It is clear from \cref{eq:updating} that the set of null distributions $\PP_0$ is responsible for the specification of \mesostructure{} assignments in the absence of interlayer dependencies (i.e., if $P_{i,\balpha}^{j,\bbeta} = 0$ for all $(i,\balpha)$, $(j,\bbeta)$). In general, $\hat{p}_{j,\bbeta}$ determines the relative importance of interlayer dependencies and the null distribution on the \mesostructure{} assignment of state node $(j,\bbeta)$. Specifically, when $\hat{p}_{j,\bbeta} =0$, the \mesostructure{} assignment of $(j,\bbeta)$ depends only on the null distribution; and when $\hat{p}_{j,\bbeta} =1$, the \mesostructure{} assignment of $(j,\bbeta)$ depends only on the \mesostructure{} assignments of its in-neighbors in the interlayer-dependency network. 

\subsubsection*{Update order}
The order in which we update \mesostructure{} assignments of state nodes via \cref{eq:updating} can influence a generated multilayer partition.
As we mentioned in \cref{sec:multilayer_networks}, an aspect of a multilayer network can be either ordered or unordered, and the update order is of particular importance when generating a multilayer partition with at least one ordered aspect. In particular, the structure of the interlayer-dependency tensor should reflect the causality that arises from the order of the aspect's elements. For an ordered aspect, structure in a given layer should depend directly only on structure in previous layers. Formally, for each ordered aspect~$a\in\set{O}$ of a multilayer network, we require that 
\begin{equation}\label{eq:ordered}
	{P_{i,\balpha}^{j,\bbeta} > 0} \Rightarrow {\alpha_a \leq \beta_a}  \, ,
\end{equation}
where $\alpha_a$ denotes the element of state $\balpha$ corresponding to aspect~$a$ and where (as we stated in \cref{sec:notation}) we require that the labels of an aspect's elements reflect the ordering of those elements~\footnote{For a fully-ordered multilayer network (i.e., where $\set{U}=\emptyset$), condition~\eqref{eq:ordered} is equivalent to the existence of a matrix representation of the flattened interlayer-dependency tensor that is upper triangular. (See \cref{sfig:temporal} for an example.) In the case of a partially-ordered multilayer network, the same equivalence holds if we \quoting{aggregate}~\cite{DeDomenico2013} the interlayer-dependency tensor over unordered aspects.}. For example, if we think of the interlayer edges in \cref{fig:notation} as interlayer edges in the interlayer-dependency network, then \cref{eq:ordered} states that all edges with nonzero edge weights must respect the arrow of time. That is, it is impossible to have edges 
of the form $((j, (2, \alpha_2)), (i, (1, \alpha_2)))$, with nodes $i,j\in\{1,2,3\}$ and aspect $\alpha_2\in\{1,2,3\}$.

The update order for the state nodes also needs to reflect the causality from the ordered aspects. In particular, if ${a \in \set{O}}$ is an ordered aspect and we update a state node in a layer $\bbeta$, then the final \mesostructure{} assignment for state nodes in layers $\balpha$ with $\alpha_a < \beta_a$ should already be fixed. For example, in \cref{fig:notation}, an edge direction should respect the arrow of time; and it is also necessary that the assignment of all state nodes in the first time point are fixed before one updates the assignments of state nodes in the second time point. To achieve the latter, we divide the layers into sets of \todefine{order-equivalent} layers, such that two layers $\balpha$ and $\bbeta$ are \define{order-equivalent} if and only if ${\alpha_a=\beta_a}$ for all ${a\in\set{O}}$. We also say that layer $\balpha$ \define{precedes} layer $\bbeta$ if and only if ${\alpha_a<\beta_a}$ for all ${a \in \set{O}}$. (For example, the three layers in \cref{fig:notation} that correspond to the first time point are order-equivalent and precede the three layers that correspond to the second time point.) Based on this equivalence relation, we obtain an ordered set of equivalence classes by inheriting the order from the ordered aspects. If we sort the classes of order-equivalent layers based on \todefine{lexicographic} order~\cite{note_lexi} of the ordered aspects, then (as a consequence of \cref{eq:ordered}) the \mesostructure{} assignment of state nodes in a given layer depends only on the assignment of state nodes in order-equivalent or preceding layers. This ensures that our partition-generation process reflects notions of causality that arise in multilayer networks with at least one ordered aspect. 

\subsubsection*{Sampling process}

The general idea of our sampling process is to simultaneously sample the \mesostructure{} assignment of state nodes within each class of order-equivalent layers and to sequentially sample the \mesostructure{} assignment of state nodes in non-order-equivalent layers, conditional on the fixed assignment of state nodes in preceding layers.

More specifically, our sampling algorithm proceeds as follows. First, we sample an initial multilayer partition from the null distribution (i.e., $S_{i,\balpha}(0) \sim \PP_0^{\balpha}$). Second,  we sample a partition for the first class of order-equivalent layers using (pseudo-)Gibbs sampling. In particular, we iteratively sample a layer uniformly at random from the first class of order-equivalent layers and update the \mesostructure{} assignment of all state nodes in that layer~\footnote{The order in which we update state nodes in the same layer has no effect on the sampling process, because their conditional distributions are independent. One can even update them in parallel.} based on \cref{eq:updating}. This defines a Markov chain on a subspace of multilayer partitions. We repeat the update process for sufficiently many iterations such that we approximately sample the \mesostructure{} assignments of the first class of order-equivalent layers from a stationary distribution of this Markov chain. Third, we sample state node assignments for subsequent classes of order-equivalent layers in the same way, based on a fixed state-node assignment from preceding layers. In particular, at each update step $\tau$ in \cref{eq:updating}, a state node can either copy a fixed \mesostructure{} assignment from an in-neighbor in a preceding layer, copy a current \mesostructure{} assignment from an in-neighbor in an order-equivalent layer, or obtain a \mesostructure{} assignment from the null distribution. 

For a fully-ordered multilayer network (i.e., a multilayer network with ${\set{U}=\emptyset}$), each class of order-equivalent layers consists only of a single layer. Consequently, one needs only a single update for each state node for the sampling process to converge. (Subsequent updates would constitute independent samples from the same distribution.) For an unordered multilayer network, our sampling algorithm reduces to (pseudo-)Gibbs sampling, because all layers are order-equivalent and there is thus only one class of order-equivalent layers. For a partially-ordered multilayer network, there are multiple non-singleton classes of order-equivalent layers. We use (pseudo-)Gibbs sampling on each class, conditional on the \mesostructure{} assignment of state nodes in preceding layers. 

In \cref{app:sampling}, we explain our sampling process in more detail. We describe our (pseudo-)Gibbs sampling procedure for each class of order-equivalent layers, and we show pseudo-code that one can use to sample partitions in unordered, partially-ordered, and fully-ordered multilayer networks in \cref{alg:sample_partition}. We also discuss the convergence properties of our sampling procedure for unordered and partially-ordered multilayer networks, including the effect of the potential incompatibility of the distributions that are defined by \cref{eq:updating}.

%%%%%%%%%%%%%%%%%%%%%%%%%%%%%%%%
%%%%%%%%%%%%%%%%%%%%%%%%%%%%%%%%

\subsection{Layer-coupled and diagonal interlayer dependencies}\label{lessgeneral}

As we mentioned at the beginning of \cref{sec:sampling_communities}, a particularly useful restriction of the interlayer-dependency tensor that still allows us to represent many situations of interest is to assume that it is layer-coupled and diagonal. For simplicity, we also assume that the multilayer network is fully interconnected. Under these assumptions, we can express the interlayer-dependency tensor $\mat{P}$ using a \define{layer-coupled interlayer-dependency tensor} $\mat{\LP}$ with elements $P_{i,\balpha}^{j,\bbeta}=\delta(i,j)\LP_{\balpha}^{\bbeta}$.
The probability that state node $(j,\bbeta)$ copies its \mesostructure{} assignment from an arbitrary state node when 
$(j,\bbeta)$'s \mesostructure{} assignment is updated is given by
\begin{equation}
	\hat{p}_{\bbeta} = \sum_{\balpha\in \set{L}} \LP_{\balpha}^{\bbeta}\,.
\label{copying_prob2} 
\end{equation}
As before, we require that $\hat{p}_{\bbeta} \leq 1$ and $ \LP_{\balpha}^{\balpha} = 0$ for all ${\balpha \in \set{L}}$.  The update equation (\cref{eq:updating}) then simplifies to 
\begin{equation}\label{eq:updating_layer}
\begin{aligned}
	&\PP[S_{j,\bbeta}(\tau+1)=s|\mat S(\tau)]\\[5pt]
	&\qquad \qquad=\smashoperator{\sum_{\balpha \in \set{L}}}\LP_{\balpha}^{\bbeta}\,\delta(S_{j,\balpha}(\tau),s)\\[5pt]
	&\qquad \qquad \qquad + \left({1-\hat{p}_{\bbeta}}\right) \PP_0^{\bbeta} [S_{j,\bbeta} = s]\,,
\end{aligned}
\end{equation}
which depends on the interlayer-dependency tensor $\mat{\LP}$ (which is now independent of state nodes) and the null distributions $\PP_0$. Each term $\LP_{\balpha}^{\bbeta}$ quantifies the extent to which an induced partition in layer $\bbeta$ depends directly on an induced partition in layer $\balpha$. By 
considering different $\mat{\LP}$, one can generate multilayer networks that correspond to several important scenarios, including temporal networks, multiplex networks, and multilayer networks with more than one aspect (e.g., with combinations of temporal and multiplex features). That is, the above restriction reduces the dimension of the interlayer-dependency tensor significantly, while still allowing one to analyze several very important situations.

\begin{figure}
\subfloat[Layer-coupled temporal dependencies
\label{sfig:temporal}]{%
\includegraphics{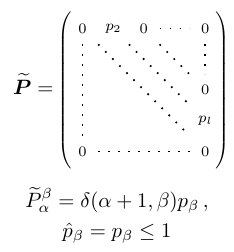}%
}%
\hfill
\subfloat[Layer-coupled multiplex dependencies \label{sfig:multiplex}]{%
\includegraphics{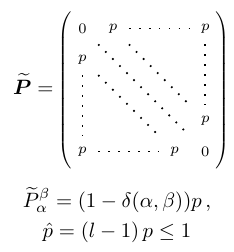}%
}

\caption{Layer-coupled interlayer dependency tensors (which, in this case, are matrices) for different types of multilayer networks with a single aspect. \sref{sfig:temporal} In a typical temporal network, an induced partition in a layer depends directly only on the induced partition in the previous layer. Therefore, the only nonzero elements of the layer-coupled interlayer dependency tensor occur in the first superdiagonal. \sref{sfig:multiplex} In a typical multiplex network, an induced partition in a layer depends directly on the induced partitions in all other layers. We show a layer-coupled example. The copying probability in \cref{copying_prob2} is $p_\beta$ for the temporal layer-coupled interlayer-dependency matrix in panel (a) and $\hat{p}$ for the multiplex layer-coupled interlayer-dependency matrix in panel (b). We suppress the subscript $\beta$ in panel (b), as the copying probability is the same for all layers.
 \label{fig:coupling_ex}}
\end{figure}

\begin{figure}
\includegraphics{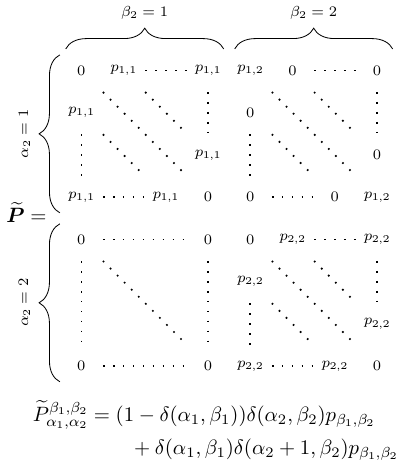}%
\caption{Block-matrix representation of a layer-coupled interlayer-dependency tensor for a multilayer network that is both temporal and multiplex. This example has more than one aspect, and it combines the features of the examples in \cref{fig:coupling_ex}. It has two classes of order-equivalent layers (where each corresponds to one time point), as indicated by the presence of two diagonal blocks. An induced partition in a layer depends directly on the induced partitions of its order-equivalent layers and on the induced partition of the corresponding layer in the preceding class of order-equivalent layers. If we think of the interlayer edges in \cref{fig:notation} as edges in an interlayer-dependency network, this matrix with diagonal blocks of size $3\times 3$  would be the corresponding layer-coupled interlayer-dependency tensor.
\label{fig:coupling_temp_mult}}
\end{figure}

In \cref{fig:coupling_ex}, we show layer-coupled interlayer-dependency tensors for two types of single-aspect multilayer networks: a temporal network and a multiplex network. As we mentioned in \cref{sec:sampling_communities}, it is useful to think of interlayer dependencies as causal links for the flow of information between layers.
 In a temporal network, it is typical to assume that an induced partition in a layer depends directly only on induced partitions in the previous layer. There are thus $l-1$ copying probabilities (one for each pair of consecutive layers), which we are free to choose. 
Common examples include choosing the same probability for each pair of consecutive layers~\cite{Ghasemian2015, Sarzynska2014} or making some of the probabilities significantly smaller than the others to introduce change points~\cite{Peel2014, Peixoto2015}. 
In a multiplex network, an induced partition in any layer can in principle depend directly on induced partitions in all other layers. This yields $l(l-1)$ copying probabilities to choose. In \cref{sfig:multiplex}, we illustrate the simplest case, in which each layer depends equally on every other layer. 
The layer-coupled interlayer-dependency tensors in \cref{fig:coupling_ex} are matrices, so we sometimes refer to them and other examples of layer-coupled interlayer-dependency tensors with a single aspect as \define{layer-coupled interlayer-dependency matrices}.

We can also generate multilayer networks with more than one aspect and can thereby combine temporal and multiplex features. In \cref{fig:coupling_temp_mult}, we illustrate how to construct a layer-coupled interlayer-dependency tensor to generate such a multilayer network on a simple example with two aspects, one of which is multiplex and the other of which is temporal.

%%%%%%%%%%%%%%%%%%%%%%%%%%%%%%%%
%%%%%%%%%%%%%%%%%%%%%%%%%%%%%%%%

\subsection{Temporal and multiplex partitions}
\label{specialcase}

\begin{figure}
\subfloat[$p=0$]{%
\includegraphics{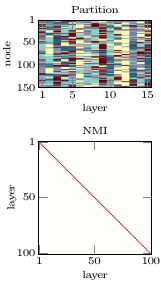}%
}%
\hfill
\subfloat[$p=0.5$]{%
\includegraphics{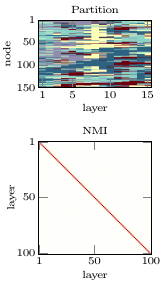}%
}%
\hfill
\subfloat[$p=0.85$]{%
\includegraphics{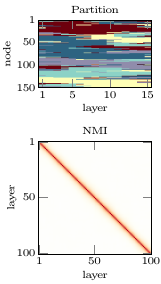}%
}\\

\subfloat[$p=0.95$]{%
\includegraphics{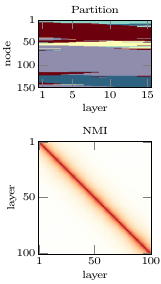}%
}%
\hfill
\subfloat[$p=0.99$]{%
\includegraphics{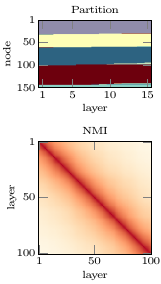}%
}%
\hfill
\subfloat[$p=1$]{%
\includegraphics{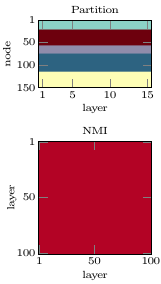}%
}%
\caption{Example temporal partitions for $(n,l) = (150,100)$. (Recall that $n$ is the number of physical nodes and $l$ is the number of layers.) 
We use the interlayer-dependency tensor from \cref{sfig:temporal} with uniform probabilities $p_\beta= p$ for all $\beta\in\{1,\ldots,l\}$ and a Dirichlet null distribution with $q=1$, $\theta = 1$, and $n_{\mathrm{set}} = 5$ (see \cref{app:nulldistribution}). 
For (a) $p = 0$, (b) $p = 0.5$, (c) $p = 0.85$, (d) $p = 0.95$, (e) $p = 0.99$, and (f) $p=1$, we show color-coded \mesostructure{} assignments ({\protect\includegraphics{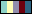}}, top) for a single example output partition and NMI values~\cite{note_nmi}  (0\,\protect\includegraphics{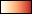}\,1, bottom) between induced partitions in different layers averaged over a sample of 10 output partitions. The parameter values match those in the numerical examples in \cref{sec:numerical_examples} (with the exception of $p=0$, which we include for completeness). We choose a node ordering for each visualization that (whenever possible) emphasizes ``persistent'' mesoscale structure~\cite{Bazzi2014}. We show only the first $15$ layers of each multilayer partition, because (as one can see in the NMI heat maps) similarities between induced partitions  for $p<1$ decay steeply with the number of layers when dependencies exist only between contiguous layers.\label{fig:temporalpartitions_ex}}
\end{figure}

\begin{figure}
\subfloat[$\hat{p}=0$]{
\includegraphics{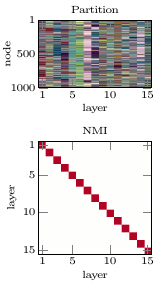}%
}%
\hfill
\subfloat[$\hat{p}=0.5$]{
\includegraphics{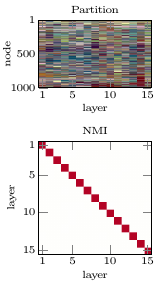}%
}%
\hfill
\subfloat[$\hat{p}=0.85$]{
\includegraphics{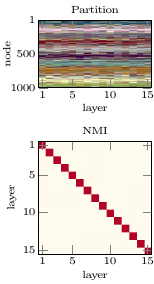}%
}%

\subfloat[$\hat{p}=0.95$]{
\includegraphics{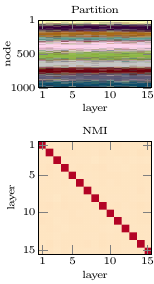}%
}%
\hfill
\subfloat[$\hat{p}=0.99$]{
\includegraphics{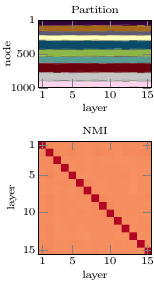}%
}%
\hfill
\subfloat[$\hat{p}=1$]{
\includegraphics{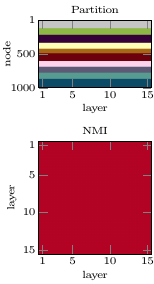}%
}

\caption{Example multiplex partitions for $(n,l) = (1000,15)$. We use the interlayer-dependency tensor in \cref{sfig:multiplex} and a Dirichlet null distribution with $q=1$, $\theta = 1$, and $n_{\mathrm{set}} = 10$ (see \cref{app:nulldistribution}). We perform $200$ updating iterations (see \cref{app:sampling}). For (a) $\hat{p} = 0$, (b) $\hat{p} = 0.5$, (c) $\hat{p} = 0.85$, (d) $\hat{p} = 0.95$, (e) $\hat{p} = 0.99$, and (f) $\hat{p}=1$, where $\hat{p} = (l-1)p$ is the probability that a state node copies its assignment from another state node, we show color-coded \mesostructure{} assignments ({\protect\includegraphics{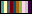}}, top) for a single example output partition and NMI
values~\cite{note_nmi} (0\,\protect\includegraphics{fig/NMI_example_colorbar_compact}\,1, bottom) between induced partitions in different layers averaged over a sample of 10 output partitions. (For the temporal case in \cref{fig:temporalpartitions_ex}, note that $\hat{p} = p$.) The parameter values match those in the numerical examples in \cref{sec:numerical_examples} (with the exception of $\hat{p}=0$, which we include for completeness). We choose a node ordering that (whenever possible) emphasizes ``persistent'' mesoscale structure~\cite{Bazzi2014}.\label{fig:multiplexpartitions_ex}}
\end{figure}

In \cref{fig:temporalpartitions_ex} and \cref{fig:multiplexpartitions_ex}, we show example multilayer partitions that we obtain with the interlayer-dependency tensors of \cref{fig:coupling_ex}. The examples in \cref{fig:temporalpartitions_ex} are temporal, and the examples in \cref{fig:multiplexpartitions_ex} are multiplex. For simplicity, we assume in \cref{sfig:temporal} that dependencies between contiguous layers are uniform  (i.e., $p_\beta = p \in[0,1]$ for all $\beta\in\{2,\ldots,l\}$). To illustrate the effect of the interlayer dependencies, we also show a heat map of the normalized mutual information (NMI)~\cite{note_nmi} between induced partitions in different layers. NMI is a measure of similarity between partitions, so we expect the values of NMI to reflect the planted dependencies. As expected, for the temporal examples, partitions are most similar for adjacent layers. Similarities decay with the distance between layers and increase with the value of $p$. For the multiplex examples, we obtain approximately uniform similarities between pairs of layers (for all pairs of layers), where the similarities increase with the value of $\hat{p}$. We use the examples of interlayer dependency of \cref{fig:temporalpartitions_ex,fig:multiplexpartitions_ex}, as well as non-uniform and multi-aspect examples, in our numerical experiments of \cref{sec:numerical_examples}.

To provide a detailed illustration of the steps that result in a multilayer partition, we focus on the temporal examples 
in \cref{fig:temporalpartitions_ex}. For the important special case of temporal interlayer dependencies,
the generative model for multilayer partitions simplifies significantly. In particular, there is a single ordered aspect, so the layer index $\alpha\in\N$ is a scalar and the order of the layers corresponds to temporal ordering. Furthermore, as we mentioned in \cref{generalcase}, for a fully-ordered multilayer network, we require that the order of the \mesostructure{}-assignment update process in \cref{eq:updating} respects the order of the layers. The update order of state nodes $(1,\alpha) \ldots (n,\alpha)$ in any given layer $\alpha$ can be arbitrary (so we can update them simultaneously), but each update is conditional on the \mesostructure{} assignment of state nodes in layer $\alpha -1$.

The update process that we described in \cref{generalcase} reduces to three steps: (1) initialize (in an arbitrary order) the \mesostructure{} assignments in layer $\alpha = 1$; (2) take the \mesostructure{} assignment of $(i,{\alpha+1})$ to be that of $(i,{\alpha})$ with probability $p$, and sample the \mesostructure{} assignment of  $(i,{\alpha+1})$ from $\mathbb{P}_0^{\alpha+1}$ with complementary probability $1-p$; and (3) repeat steps (1) and (2) until $\alpha+1=l$. As we mentioned in \cref{generalcase}, convergence is not an issue for this case (or, more generally, for any fully-ordered multilayer network), as we  need only one iteration through the layers. This three-step generative model for temporal partitions was also suggested by Ghasemian et al~\cite{Ghasemian2018}. They used it to derive a detectability threshold for the case in which the null distributions are uniform across \mesostructure{}s (i.e., $\theta \to \infty$ in \cref{nulldistribution}) and intralayer edges are generated independently using the standard SBM. In other words, one replaces the degree-corrected SBM in \cref{sec:sampling_networks} with the non-degree-corrected SBM of \cite{Holland1983}. In \cref{app:temporal}, we highlight properties of a sampled temporal partition that illustrate how the interplay between $p$ and the null distributions affects the evolution of \mesostructure{}s across layers (e.g., growth, birth, and death of induced \mesostructure{}s). The properties that we highlight hold for any choice of null distribution. In the same appendix, we subsequently illustrate that the particular choice of null distributions can greatly influence resulting partitions. For example, a non-empty overlap between the supports of null distributions that correspond to contiguous layers is a necessary condition for \mesostructure{}s to gain new state nodes over time.  The properties and observations in \cref{app:temporal} are independent of one's choice of planted-partition network model.

%%%%%%%%%%%%%%%%%%%%%%%%%%%%%%%%
%%%%%%%%%%%%%%%%%%%%%%%%%%%%%%%%

\section{Generating network edges}
\label{sec:sampling_networks}

There are diverse different types of multilayer networks \cite{Kivela2014, Porter2018}. One common type of empirical multilayer network to study is one with only intralayer edges. There are also many empirical multilayer networks with both intralayer and interlayer edges (e.g., multimodal transportation networks~\cite{gallotti2014anatomy}), as well as ones with only interlayer edges (e.g., temporal networks with edge delays, such as departures and arrivals of flights between airports~\cite{Boccaletti2014, Zanin2013}). One can use our framework to generate all three types of examples, provided the underlying edge generation model is a planted-partition network model.

Having generated a multilayer partition $\set{S}$ with dependencies between induced partitions in different layers, the simplest way to generate edges is to use any single-layer planted-partition network model (e.g., SBMs~\cite{Karrer2011, Holland1983, Jacobs2014, Peixoto2018, Aicher2014, Galhotra2017}, models for spatially-embedded networks~\cite{Sarzynska2014,Newman2015}, and so on) and to generate edges for each layer independently.
This yields a multilayer network with only intralayer edges, such that any dependencies between different layers result only from dependencies between the partitions that are induced on the different layers.  For our numerical experiments in \cref{sec:numerical_examples}, we use a single-layer network model that is a slight variant (avoiding the creation of self-edges and multi-edges) of the degree-corrected SBM (DCSBM) benchmark from~\cite{Karrer2011}, where \quoting{DCSBM benchmark} designates the specific type of DCSBM that was used in the numerical experiments of~\cite{Karrer2011}. We include pseudocode for our implementation of the DCSBM benchmark in \cref{alg:dcsbm} of \cref{app:MDCSBM}.  

One can also include dependencies between layers other than those that are induced by planted mesoscale structures. For example, one can introduce dependencies between the parameters of a single-layer planted-partition 
network model by (1) sampling them from a common probability distribution (e.g., interlayer degree correlations~\cite{Kivela2014, Lee2012} in a DCSBM) or by (2) introducing interlayer edge correlations, given a single-layer partition on each layer~\cite{Pamfil2019}. For temporal networks, one can also incorporate \quoting{burstiness} \cite{Holme2012,Holme2015,Lambiotte2013} in the inter-event-time distribution of edges. In such a scenario, the probability for an edge to exist in a given layer depends not only on the induced partition on that layer but also on the existence of the edge in previous layers. For example, one can use a Hawkes process to specify the time points at which edges are active \cite{Daley2003, Hawkes1971}.

In \cref{app:MDCSBM}, we describe a generalization of the DCSBM in~\cite{Karrer2011} to multilayer networks that one can use to generate intralayer edges and/or interlayer edges. Our generalization constitutes a framework within which we formulate the parameters of a single-layer DCSBM in a multilayer setting. One can use our multilayer DCSBM (M-DCSBM) framework to incorporate some of the features that we described in the previous paragraph (e.g., degree correlations) in a multilayer network with intralayer and/or interlayer edges.

%%%%%%%%%%%%%%%%%%%%%%%%%%%%%%%%
%%%%%%%%%%%%%%%%%%%%%%%%%%%%%%%%

\section{Numerical examples}
\label{sec:numerical_examples}

In this section, we use our framework to construct benchmark models for multilayer community-detection methods and algorithms. We use the examples of interlayer-dependency tensors from \cref{fig:coupling_ex,fig:coupling_temp_mult} to generate benchmark models. We also consider a variant of \cref{sfig:multiplex} in which we split the layers into groups, use uniform dependencies between layers in the same groups, and treat layers in different groups as independent of each other. These examples cover commonly studied temporal and multiplex dependencies, and they illustrate how one can generate benchmark multilayer networks with more than one aspect. We focus on a couple of popular quality functions for multilayer community detection in our numerical examples, rather than investigating any given method or algorithm in detail. 

We compare the behavior of several variants of \cite{genlouvain}, which are similar to the the locally greedy Louvain computational heuristic~\cite{Blondel2008}, to optimize a multilayer modularity objective function~\cite{Mucha2010,Bazzi2014} using the standard Newman--Girvan null model (which is a variant of a \quoting{configuration model} \cite{Fosdick2016}). Modularity is an objective function that is often used to partition sets of nodes into communities that have a larger total internal edge weight than the expected total internal edge weight in the same sets in a \quoting{null network}~\cite{Bazzi2014}, which is generated from some null model. Modularity maximization consists of finding a partition that maximizes this difference. For our numerical experiments, we use the generalization of modularity to multilayer networks of~\cite{Mucha2010}. For multilayer modularity, the strength of the interactions between different layers are governed by an interlayer-coupling tensor that controls the incentive for state nodes in different layers to be assigned to the same community. We use multilayer modularity with uniform interlayer coupling, so the strength of the interactions between different layers of the network depends on a layer-independent and node-independent interlayer coupling weight $\omega\geq 0$. We use diagonal and categorical (i.e., between all pairs of layers) interlayer coupling with weight $\omega$ for the multiplex examples in \cref{sec:multiplex_examples}. We use diagonal and ordinal (i.e., between contiguous layers) interlayer coupling with weight $\omega$ for the temporal examples in \cref{sec:temporal_examples}. In \cref{app:louvain}, we describe the Louvain algorithm and the variants of it that we use in this section.

We compare the results of multilayer modularity maximization with those of multilayer \textsc{InfoMap}~\footnote{We use version 0.18.2 of the \textsc{InfoMap} code with the ``\texttt{--two-level}'' option. The code is available at \url{http://mapequation.org/code}.}~\cite{DeDomenico2014a}, which uses an objective function called the ``map equation'' (which is not an equation), based on a discrete-time random walk and ideas from coding theory, to coarse-grain sets of nodes into communities \cite{Rosvall2007}. In multilayer \textsc{InfoMap}, one uses a probability $r\in[0,1]$ called the ``relaxation rate'' to control the relative frequency with which a random walker remains in the same layer or moves to other layers. (A random walker cannot change layers when $r=0$.) The relaxation rate thus controls the interactions between different layers of a multilayer network. We allow the random walker to move to all other layers when $r\neq 0$ for the multiplex examples in \cref{sec:multiplex_examples}, and we allow the random walker to move only to adjacent layers for the temporal examples in \cref{sec:temporal_examples}. In contrast to multilayer modularity (where $\omega=0$ yields single-layer modularity), single-layer $\textsc{InfoMap}$ is not equivalent to choosing $r=0$ in multilayer $\textsc{InfoMap}$ because placing state nodes that correspond to the same physical node in the same community can contribute positively to the quality function even when $r=0$~\cite{DeDomenico2014a}. Consequently, we compute single-layer $\textsc{InfoMap}$ separately and reference it by the data point ``s" on the horizontal axis.

In all experiments in this section, we generate a multilayer partition using our copying process in \cref{sec:sampling_communities}  (see \cref{alg:sample_partition} in \cref{app:sampling}) and a multilayer network for a fixed planted partition using a slight variant of the DCSBM benchmark network model in~\cite{Karrer2011} that avoids generating self-edges and multi-edges (see \cref{alg:dcsbm} in \cref{app:MDCSBM}). 
This produces multilayer networks that have only intralayer edges and in which the connectivity patterns in different layers depend on each other.
Following~\cite{Karrer2011}, we parametrize the DCSBM benchmark in terms of its distribution of expected degrees and a community-mixing parameter $\mu\in[0,1]$ that controls the strength of the community structure in the sampled network edges. For $\mu=0$, all edges lie within communities; for $\mu=1$, edges are distributed independently of the communities, where the probability of observing an edge between two state nodes in the same layer depends only on the expected degrees of those two state nodes. We use a truncated power law for the distribution of expected degrees.

The DCSBM benchmark imposes community structure as an expected feature of an ensemble of networks that it generates. 
The definition of the mixing parameter $\mu$ of the DCSBM benchmark ensures that the planted partition remains community-like --- as intra-community edges are more likely to be observed and inter-community edges are less likely to be observed than in a network that is generated from a single-block DCSBM with the same expected degrees --- for any value of $\mu<1$. Consequently, given sufficiently many samples from the same DCSBM benchmark (i.e., all samples have the same planted partition and same expected degrees), one should be able to identify the planted community structure for any value of $\mu<1$ (where the necessary number of samples goes to infinity as $\mu \to 1$). This feature makes the DCSBM benchmark an interesting test for the ability of multilayer community-detection methods to aggregate information from multiple layers. 

We use NMI~\cite{note_nmi} to compare the performance of different community-detection algorithms. For each of our partitions,
we compute the mean of the NMI between the partition induced on each layer by the output partition and that induced by the planted partition. That is,
\begin{equation*}
	\mean{\NMI}(\set{S}, \set{T}) = \frac{1}{l}\sum_{\balpha \in \set{L}} \NMI(\set{S}|_\balpha, \set{T}|_\balpha)\,.
\end{equation*} 
The quantity $\mean{\NMI}$ is invariant under permutations of the \mesostructure{} labels within a layer. Consequently, $\mean{\NMI}$ is well-suited to comparing multilayer community-detection methods with single-layer community-detection methods. In particular, it allows us to test whether multilayer community-detection methods can exploit dependencies between layers of a multilayer network when $\hat{p}\gg 0$ (see \cref{copying_prob2}). In \cref{app:mNMI}, we show numerical experiments in which we compute NMI between multilayer partitions. We denote the NMI between two multilayer partitions by mNMI.

\begin{figure*}
\includegraphics{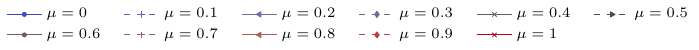}\\

\begin{tabular}{@{}p{\linewidth}@{}}
\textsc{GenLouvain}\\
\hline
\vspace{-1.5em}
\subfloat[$\hat{p}=0.5$\label{sfig:M_L_0.5}]{
\includegraphics{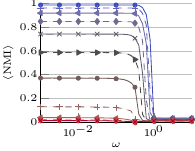}%
}%
\hfill
\subfloat[$\hat{p}=0.85$\label{sfig:M_L_0.85}]{
\includegraphics{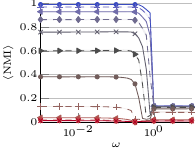}%
}%
\hfill
\subfloat[$\hat{p}=0.95$\label{sfig:M_L_0.95}]{
\includegraphics{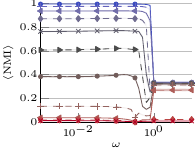}%
}%
\hfill
\subfloat[$\hat{p}=0.99$\label{sfig:M_L_0.99}]{
\includegraphics{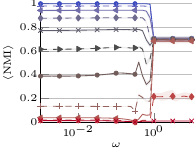}%
}%
\hfill
\subfloat[$\hat{p}=1$\label{sfig:M_L_1}]{
\includegraphics{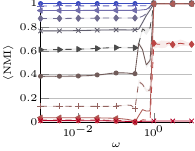}%
}\\
\vspace{0em}
\textsc{GenLouvainRand}\\
\hline
\vspace{-1.5em}
\subfloat[$\hat{p}=0.5$\label{sfig:M_LRW_0.5}]{
\includegraphics{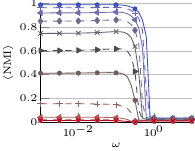}%
}%
\hfill
\subfloat[$\hat{p}=0.85$\label{sfig:M_LRW_0.85}]{
\includegraphics{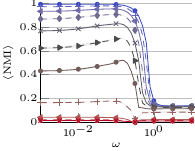}%
}%
\hfill
\subfloat[$\hat{p}=0.95$\label{sfig:M_LRW_0.95}]{
\includegraphics{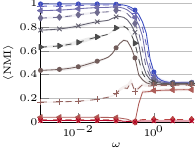}%
}%
\hfill
\subfloat[$\hat{p}=0.99$\label{sfig:M_LRW_0.99}]{
\includegraphics{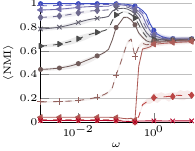}%
}%
\hfill
\subfloat[$\hat{p}=1$\label{sfig:M_LRW_1}]{
\includegraphics{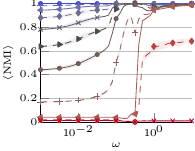}%
}\\
\vspace{0em}
\textsc{InfoMap}\\
\hline
\vspace{-1.5em}
\subfloat[$\hat{p}=0.5$\label{sfig:M_Inf_0.5}]{
\includegraphics{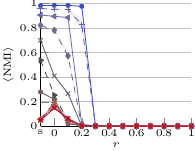}%
}%
\hfill
\subfloat[$\hat{p}=0.85$\label{sfig:M_Inf_0.85}]{
\includegraphics{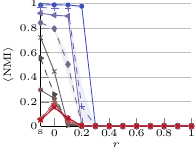}%
}%
\hfill
\subfloat[$\hat{p}=0.95$\label{sfig:M_Inf_0.95}]{
\includegraphics{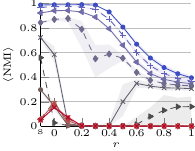}%
}%
\hfill
\subfloat[$\hat{p}=0.99$\label{sfig:M_Inf_0.99}]{
\includegraphics{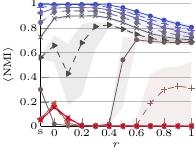}%
}%
\hfill
\subfloat[$\hat{p}=1$\label{sfig:M_Inf_1}]{
\includegraphics{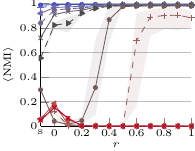}%
}%
\end{tabular}
\caption{
{Multiplex networks with uniform interlayer dependencies.} Effect of interlayer coupling strength $\omega$ and relaxation rate $r$ on the ability of different community-detection algorithms to recover planted partitions as a function of the community-mixing parameter $\mu$ in benchmark networks with uniform multiplex dependencies (see \cref{sfig:multiplex}). Each multilayer network has $1000$ nodes and $15$ layers, and each node is present in all layers.  All NMI values are means over $10$ runs of the algorithms and $100$ instantiations of the benchmarks. (See the introduction of \cref{sec:numerical_examples}.) Each curve corresponds to the mean NMI values that we obtain for a given value of $\mu$, and the shaded area around a 
curve corresponds to the minimum and maximum NMI values that we obtain with the $10$ sample partitions for a given value of $\hat{p}$. 
\label{fig:mult-bench}}
\end{figure*}

\begin{figure*}[ht!]
\centering
\includegraphics{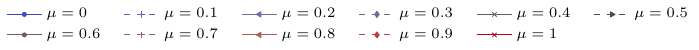}\\
\begin{tabular}{@{}p{\linewidth}@{}}
\textsc{GenLouvain}\\
\hline\vspace{-1.5em}
\subfloat[$p=0.5$\label{sfig:T_L_0.5}]{
\includegraphics{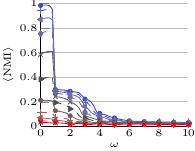}%
}%
\hfill
\subfloat[$p=0.85$\label{sfig:T_L_0.85}]{
\includegraphics{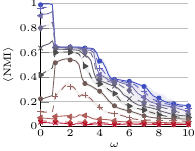}%
}%
\hfill
\subfloat[$p=0.95$\label{sfig:T_L_0.95}]{
\includegraphics{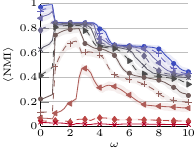}%
}%
\hfill
\subfloat[$p=0.99$\label{sfig:T_L_0.99}]{
\includegraphics{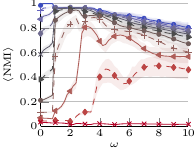}%
}%
\hfill
\subfloat[$p=1$\label{sfig:T_L_1}]{
\includegraphics{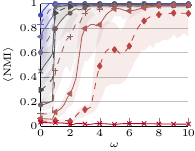}%
}\\
\vspace{0em}
\textsc{GenLouvainRand}\\
\hline\vspace{-1.5em}
\subfloat[$p=0.5$\label{sfig:T_LRW_0.5}]{
\includegraphics{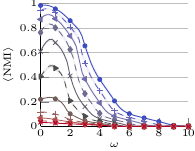}%
}%
\hfill
\subfloat[$p=0.85$\label{sfig:T_LRW_0.85}]{
\includegraphics{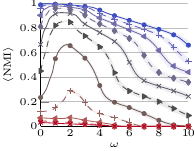}%
}%
\hfill
\subfloat[$p=0.95$\label{sfig:T_LRW_0.95}]{
\includegraphics{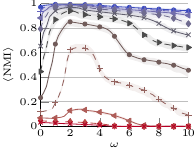}%
}%
\hfill
\subfloat[$p=0.99$\label{sfig:T_LRW_0.99}]{
\includegraphics{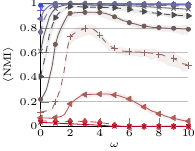}%
}%
\hfill
\subfloat[$p=1$\label{sfig:T_LRW_1}]{
\includegraphics{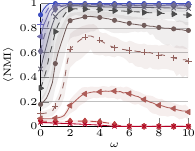}%
}\\
\vspace{0em}
\textsc{InfoMap}\\
\hline\vspace{-1.5em}
\subfloat[$p=0.5$\label{sfig:T_Inf_0.5}]{
\includegraphics{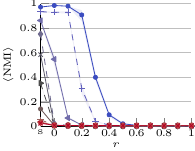}%
}%
\hfill
\subfloat[$p=0.85$\label{sfig:T_Inf_0.85}]{
\includegraphics{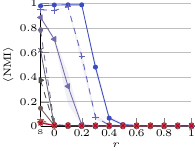}%
}%
\hfill
\subfloat[$p=0.95$\label{sfig:T_Inf_0.95}]{
\includegraphics{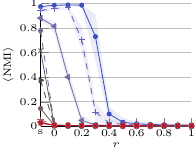}%
}%
\hfill
\subfloat[$p=0.99$\label{sfig:T_Inf_0.99}]{
\includegraphics{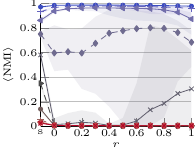}%
}%
\hfill
\subfloat[$p=1$\label{sfig:T_Inf_1}]{
\includegraphics{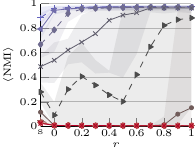}%
}\\
\end{tabular}
\caption{Temporal networks with uniform interlayer dependencies. Effect of interlayer coupling strength $\omega$ and relaxation rate $r$ on the ability of different community-detection algorithms to recover planted partitions as a function of the community-mixing parameter $\mu$ in a temporal benchmark with uniform interlayer dependencies (i.e., $p_\beta = p \in[0,1]$ for all $\beta\in\{2,\ldots,l\}$ in \cref{sfig:temporal}). Each multilayer network has $150$ nodes and $100$ layers, and each node is present in all layers.  All NMI values are means over $10$ runs of the algorithms and $100$ instantiations of the benchmark. (See the introduction of \cref{sec:numerical_examples}.) Each curve
corresponds to the mean NMI values that we obtain for a given value of $\mu$, and the shaded area around a curve
corresponds to the minimum and maximum NMI values that we obtain with the $10$ sample partitions for a given value of $p$. 
}
\label{fig:temp-bench}
\end{figure*}

In all of our numerical experiments in \cref{sec:multiplex_examples,sec:temporal_examples}, we sample the benchmark networks in the following way. For each value of $\hat{p}$, we generate $10$ sample partitions. For each sample partition and value of $\mu$, we generate $10$ sample multilayer networks. This yields $100$ 
benchmark instantiations for each pair $(\hat{p}, \mu)$. We run each community-detection algorithm $10$ times on each 
instantiation. In \cref{fig:mult-bench,fig:temp-bench,fig:Mhet-bench,fig:tempCP-bench}, we show $\mean{\mathrm{NMI}}$ between planted and recovered multilayer partitions {averaged over sample partitions, sample networks, and algorithmic runs for each value of $\mu$. The results for different planted partitions are generally similar. 
The only exception is \textsc{InfoMap} on temporal networks with $p=0.99$ and $p=1$, where we observe large differences between results for different partitions for certain values of $\mu$.}

\Cref{fig:mult-bench,fig:temp-bench,fig:Mhet-bench,fig:tempCP-bench} correspond to different choices for the interlayer-dependency tensor. In each figure, rows correspond to different choices for the community-detection algorithm, and columns correspond to different values of the copying probabilities (with the strength of interlayer dependencies increasing from left to right). All benchmark instantiations that we use for  
\cref{fig:mult-bench,fig:temp-bench,fig:Mhet-bench,fig:tempCP-bench,fig:multiaspect_example} are available at \url{https://dx.doi.org/10.5281/zenodo.3304059}.

%%%%%%%%%%%%%%%%%%%%%%%%%%%%%%%%
%%%%%%%%%%%%%%%%%%%%%%%%%%%%%%%%

\subsection{Multiplex examples}
\label{sec:multiplex_examples}

\begin{figure}
\centering
\includegraphics{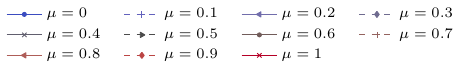}\\
\begin{tabular}{@{}p{\linewidth}@{}}
\textsc{GenLouvain}\\
\hline
\vspace{-1.5em}
\subfloat[$\hat{p}=0.85$, $p_c = 0$\label{sfig:Mhet_L_0.85}]{
\includegraphics{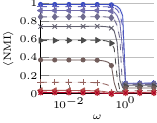}%
}%
\hfill
\subfloat[$\hat{p}=0.95$, $p_c = 0$\label{sfig:Mhet_L_0.95}]{
\includegraphics{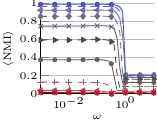}%
}%
\hfill
\subfloat[$\hat{p}=1$, $p_c = 0$\label{sfig:Mhet_L_1}]{
\includegraphics{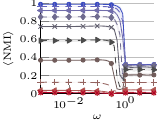}%
}\\
\vspace{0em}
\textsc{GenLouvainRand}\\
\hline
\vspace{-1.5em}
\subfloat[$\hat{p}=0.85$, $p_c = 0$\label{sfig:Mhet_LRW_0.85}]{
\includegraphics{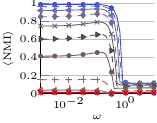}%
}%
\hfill
\subfloat[$\hat{p}=0.95$, $p_c = 0$\label{sfig:Mhet_LRW_0.95}]{
\includegraphics{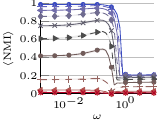}%
}%
\hfill
\subfloat[$\hat{p}=1$, $p_c = 0$\label{sfig:Mhet_LRW_1}]{
\includegraphics{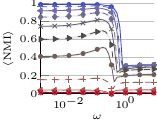}%
}\\
\vspace{0em}
\textsc{InfoMap}\\
\hline
\vspace{-1.5em}
\subfloat[$\hat{p}=0.85$, $p_c = 0$\label{sfig:Mhet_Inf_0.85}]{
\includegraphics{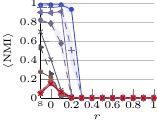}%
}%
\hfill
\subfloat[$\hat{p}=0.95$, $p_c = 0$\label{sfig:Mhet_Inf_0.95}]{
\includegraphics{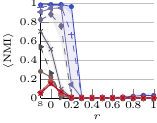}%
}%
\hfill
\subfloat[$\hat{p}=1$, $p_c = 0$\label{sfig:Mhet_Inf_1}]{
\includegraphics{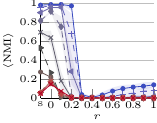}%
}%
\end{tabular}
\caption{Multiplex networks with nonuniform interlayer dependencies. Effect of interlayer coupling strength $\omega$ and relaxation rate $r$ on the ability of different community-detection algorithms to recover planted partitions as a function of the community-mixing parameter $\mu$ in a multiplex benchmark with nonuniform interlayer dependencies.  Each multilayer network has $1000$ nodes and $15$ layers, and each node is present in all layers. The layer-coupled interlayer-dependency matrix is a block-diagonal matrix with diagonal blocks of size $5\times 5$. For each diagonal block, we set the value in the interlayer-dependency matrix to a value $p$ (so each diagonal block has the same structure as the matrix in \cref{sfig:multiplex}); for each off-diagonal block, we set the value to $p_c = 0$, thereby incorporating
an abrupt change in community structure.  All NMI values are means over $10$ runs of the algorithms and $100$ instantiations of the benchmarks. 
Each curve corresponds to the mean NMI values that we obtain for a given value of $\mu$, and the shaded area around a curve corresponds to the minimum and maximum NMI values that we obtain with the $10$ sample partitions for a given value of $\hat{p}$.
\label{fig:Mhet-bench}}
\end{figure}

In this section, we consider two stylized examples of multiplex networks. Multiplex networks arise in a variety of different applications, including international relations and trade~\cite{Cranmer:2014ut,Barigozzi2010,Mastrandrea2014}, social networks~\cite{Fan2015}, and ecological networks~\cite{Kefi2016}. In our multiplex examples, we consider simple dependency structures in which we expect multilayer community-detection methods to outperform single-layer methods by exploiting interlayer dependencies.

In \cref{fig:mult-bench}, we consider multiplex networks with uniform dependencies between community structure in different layers. In \cref{fig:Mhet-bench}, we consider multiplex networks with nonuniform dependencies between community structure in different layers. In both figures, we parametrize the amount of interlayer dependency in a network by the probability $\hat{p}$ (see \cref{copying_prob2}) that a state node copies its community assignment from a neighbor in the interlayer-dependency network. All multilayer networks have $n=1000$ physical nodes and $l=15$ layers. Each node is present in every layer, so there are a total of $15000$ state nodes. In \cref{sfig:multiplex}, we show the layer-coupled interlayer-dependency matrix that we use to generate the uniform multiplex networks. For the nonuniform multiplex networks, we split the layers into $3$ groups of $5$ layers each. We use uniform dependencies between layers in the same group, and layers in different groups are independent of each other. The resulting  layer-coupled interlayer-dependency matrix is block diagonal with diagonal blocks as in \cref{sfig:multiplex} and $0$ entries on the off-diagonal blocks. We use a Dirichlet null distribution with ${n_{\mathrm{set}}=10}$, ${\theta=1}$, and ${q=1}$ (see \cref{app:nulldistribution}) to specify expected community sizes and the M-DCSBM benchmark (see \cref{app:MDCSBM}) with ${\eta_k=2}$, ${k_{\mathrm{min}}=3}$, and ${k_{\mathrm{max}}=150}$ to generate intralayer edges. We perform $200$ iterations of our update process (see \cref{app:sampling}). 

Comparing our results for \textsc{GenLouvain} (\cref{sfig:M_L_0.5,sfig:M_L_0.85,sfig:M_L_0.95,sfig:M_L_0.99,sfig:M_L_1,sfig:Mhet_L_0.85,sfig:Mhet_L_0.95,sfig:Mhet_L_1}) and \textsc{GenLouvainRand} with reiteration and post-processing (\cref{sfig:M_LRW_0.5,sfig:M_LRW_0.85,sfig:M_LRW_0.95,sfig:M_LRW_0.99,sfig:M_LRW_1,sfig:Mhet_LRW_0.85,sfig:Mhet_LRW_0.95,sfig:Mhet_LRW_1}), we see that the choice of optimization heuristic for a given quality function can significantly affect the quality of resulting identified partitions. In particular, for \textsc{GenLouvain}, we observe two distinct regimes and what appears to be a sharp transition between them~\footnote{We observe some erratic behavior at the transition as the interlayer coupling $\omega$ approaches $1$ from below. For values of $\omega$ near $1$ but smaller than $1$, the \textsc{GenLouvain} algorithm has a tendency to place all state nodes in a single community. This observation is related to the transition behavior that was described in \cite{Bazzi2014}. For values of $\omega$ above a certain threshold~\cite{Bazzi2014} only interlayer merges occur in the first phase of Louvain.}. For $\omega<1$, we obtain a partition that is of similar quality to what we obtain by maximizing single-layer modularity; however, for $\omega>1$, we obtain a partition with identical induced partitions for each layer. 
We call the latter an \define{aggregate partition}. Although the aggregate partition can be more similar 
to the planted partition than the single-layer partition when $\hat{p}$ is sufficiently large (e.g., this occurs in \cref{sfig:M_L_0.99,sfig:M_L_1}), we do not observe an interval of $\omega$ values between the two regimes in which \textsc{GenLouvain} 
recovers the planted partition better than both the single-layer and aggregate partitions.
By contrast, \textsc{GenLouvainRand} with reiteration and post-processing identifies partitions that match the planted partition more closely than either the single-layer or aggregate partitions in most cases. The exceptions are uniform multiplex networks with ${\hat{p}=0.5}$ (see \cref{sfig:M_LRW_0.5}), where it is unable to exploit the weak dependencies to outperform the single-layer case; and ${\hat{p}=1}$ (\cref{sfig:M_LRW_1}), where the aggregate partition is always best. Most of the improvement in the results for \textsc{GenLouvainRand} over those for \textsc{GenLouvain} comes from reiteration. The additional randomization helps smooth out the transition at $\omega\approx 1$. Post-processing only yields a minor improvement in the value of $\mean{\NMI}$. 
However, its effect is more pronounced if we compute the mNMI between multilayer partitions (see \cref{app:mNMI}).

Multilayer \textsc{InfoMap} exhibits some problematic behavior in these benchmark experiments. Our results for multilayer \textsc{InfoMap} are noticeably worse than those that we obtain with single-layer $\textsc{InfoMap}$ (corresponding to the data point ``s" on the horizontal axis) for networks with comparatively weak community structure (specifically, ${\mu\geq 0.4}$), unless the planted partition for different layers is very similar (i.e., unless ${\hat{p}\geq 0.99}$). Furthermore, the methods that are based on multilayer modularity outperform multilayer \textsc{InfoMap} in our experiments for networks with weak community structure and similar layers (i.e., when both $\mu$ and $\hat{p}$ are close to $1$). \textsc{InfoMap} does not identify meaningful community structure in networks with $\mu > 0.7$ in any of our numerical examples, whereas \textsc{GenLouvain} and \textsc{GenLouvainRand} identify meaningful structure even for networks with very weak community structure (e.g., with $\mu=0.9$) for sufficiently large values of $\omega$.

Our results for nonuniform multiplex networks in \cref{fig:Mhet-bench} are similar to those that we showed for the uniform multiplex networks in \cref{fig:mult-bench}. In particular, we see that \textsc{GenLouvainRand} can exploit dependencies between community structure in different layers (although the improvement over single-layer modularity is less pronounced than in the uniform examples), whereas \textsc{GenLouvain} cannot. The main difference is that for $\hat{p}=1$, the aggregate partitions from \textsc{GenLouvain} and \textsc{GenLouvainRand} do not recover the planted partition for sufficiently large values of $\omega$, because the induced partitions are not identical across layers in the planted partition. In principle, multilayer \textsc{InfoMap} should have an advantage over methods based on maximizing multilayer modularity in this case study, as the former's quality function is designed to detect breaks in community structure, whereas multilayer modularity forces some persistence of community labels between any pair of layers. In practice, however, multilayer \textsc{InfoMap} correctly identifies the planted community structure only when it is particularly strong (as we show in \cref{app:mNMI}, it correctly identifies the different groups of layers for networks with $\mu\leq 0.3$); and it is outperformed by single-layer \textsc{InfoMap} for networks with $\mu\geq 0.4$. 

We suspect that at least some of the shortcomings of multilayer \textsc{InfoMap} in these experiments are due to the use of a Louvain-like optimization heuristic, rather than from flaws in \textsc{InfoMap}'s quality function. As we have seen from the results with the heuristics \textsc{GenLouvain} and \textsc{GenLouvainRand}, seemingly minor adjustments of an optimization heuristic can have large effects on the quality of the results. We are thus hopeful that similar adjustments can also improve the results for multilayer \textsc{InfoMap}.

%%%%%%%%%%%%%%%%%%%%%%%%%%%%%%%%
%%%%%%%%%%%%%%%%%%%%%%%%%%%%%%%%

\subsection{Temporal examples}
\label{sec:temporal_examples}

\begin{figure}
\centering
\includegraphics{fig/temporalCP_colorbar_L.pdf}\\
\begin{tabular}{@{}p{\linewidth}@{}}
\textsc{GenLouvain}\\
\hline
\vspace{-1.5em}
\subfloat[$p=0.85$, $p_c = 0$\label{sfig:TCP_L_0.85}]{
\includegraphics{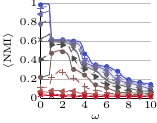}%
}%
\hfill
\subfloat[$p=0.95$, $p_c = 0$\label{sfig:TCP_L_0.95}]{
\includegraphics{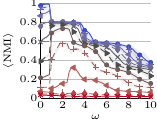}%
}%
\hfill
\subfloat[$p=1$, $p_c = 0$\label{sfig:TCP_L_1}]{
\includegraphics{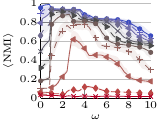}%
}\\
\vspace{0em}
\textsc{GenLouvainRand}\\
\hline
\vspace{-1.5em}
\subfloat[$p=0.85$, $p_c = 0$\label{sfig:TCP_LRW_0.85}]{
\includegraphics{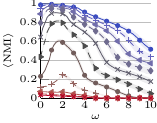}%
}%
\hfill
\subfloat[$p=0.95$, $p_c = 0$\label{sfig:TCP_LRW_0.95}]{
\includegraphics{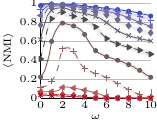}%
}%
\hfill
\subfloat[$p=1$, $p_c = 0$\label{sfig:TCP_LRW_1}]{
\includegraphics{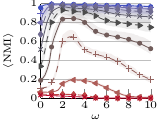}%
}\\
\vspace{0em}
\textsc{InfoMap}\\
\hline
\vspace{-1.5em}
\subfloat[$p=0.85$, $p_c = 0$\label{sfig:TCP_Inf_0.85}]{
\includegraphics{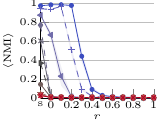}%
}%
\hfill
\subfloat[$p=0.95$, $p_c = 0$\label{sfig:TCP_Inf_0.95}]{
\includegraphics{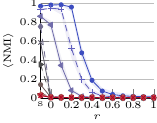}%
}%
\hfill
\subfloat[$p=1$, $p_c = 0$\label{sfig:TCP_Inf_1}]{
\includegraphics{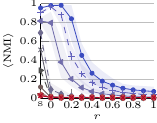}%
}%
\end{tabular}
\caption{Temporal networks with nonuniform interlayer dependencies.
Effect of interlayer coupling strength $\omega$ and relaxation rate $r$ on the ability of different community-detection algorithms to recover planted partitions as a function of the community-mixing parameter $\mu$ in a temporal benchmark with nonuniform interlayer dependencies.
Each multilayer network has $150$ nodes and $100$ layers, and each node is present in all layers. Every 25th layer (i.e., for layers 25, 50, and 75), we set the value of $p_\beta$ in \cref{sfig:temporal} to $p_c = 0$, thereby introducing an abrupt change in community structure; and we set all other values of $p_\beta$ in \cref{sfig:temporal} to $p$.  All NMI values are means over $10$ runs of the algorithms and $100$ instantiations of the benchmark.
Each curve corresponds to the mean NMI values that we obtain for a given value of $\mu$, and the shaded area around a curve corresponds to the minimum and maximum NMI values that we obtain with the $10$ sample partitions for a given value of $p$. 
}
\label{fig:tempCP-bench}
\end{figure}

Temporal networks arise in many different applications \cite{Holme2012,Holme2015}, such as the study of brain dynamics~\cite{Betzel2017}, financial-asset correlations~\cite{Bazzi2014}, and scientific citations~\cite{Hric2018}. When representing a temporal network as a multilayer network, one orders the layers in a causal way, such that structure in a particular layer depends directly only on structure in previous layers (and not in future layers). To generate multilayer networks that have such structure, we use our sampling process for fully-ordered multilayer networks and the interlayer-dependency tensor of \cref{sfig:temporal}. 

We consider two stylized examples of temporal networks. In \cref{fig:temp-bench}, we show results for temporal networks that have uniform dependencies between consecutive layers (and hence tend to evolve gradually). To generate these networks, we set ${p_\beta=p}$ for all layers in \cref{sfig:temporal}. In \cref{fig:tempCP-bench}, we show results for temporal networks with change points. To generate these networks, we set ${p_\beta=p}$ for all layers except layers 25, 50, and 75, for which ${p_\beta=p_c=0}$, resulting in abrupt changes in community structure. Each multilayer network in these two examples has $n=150$ physical nodes and $l=100$ layers. Each node is present in every layer, so there are a total of $15000$ state nodes.  We use a Dirichlet null distribution to specify expected community sizes and set $n_{\mathrm{set}} =5$, $\theta=1$, and $q=1$ (see \cref{app:nulldistribution}). For the temporal networks with change points in \cref{fig:tempCP-bench}, we choose the support of the null distribution so that communities after a change point have new labels. We use the M-DCSBM benchmark (see \cref{app:MDCSBM}) with $\eta_k= 2$, $k_{\mathrm{min}} = 3$, and $k_{\mathrm{max}} =  30$ to generate intralayer edges.

Both multilayer-modularity-based algorithms (i.e., \textsc{GenLouvain} and \textsc{GenLouvainRand}) can exploit interlayer dependencies for these temporal benchmark networks and identify partitions with significantly larger $\mean{\NMI}$ than those that we obtain with single-layer modularity (i.e., with ${\omega=0}$). Typically, the peak of $\mean{\NMI}$ seems to occur when $1\lessapprox \omega \lessapprox 4$. When ${p <1}$, one expects $\mean{\NMI}$ to decrease for sufficiently large values of $\omega$, as increasing $\omega$ further favors \quoting{persistence} \cite{Bazzi2014} in the output partition that is not present in the multilayer planted partition. 

For multilayer \textsc{InfoMap}, the results are less promising. For most parameter choices, the best result for multilayer \textsc{InfoMap} is at best similar and often worse than the result for single-layer \textsc{InfoMap}. (The $\mean{\NMI}$ value for single-layer \textsc{InfoMap} is the data point that we label with ``s'' on the horizontal axis.) An exception are the results for $p=1$ and uniform temporal dependencies (see \cref{sfig:T_Inf_1}). In this example (where induced partitions are the same across layers), increasing the value of the relaxation rate $r$ enhances the recovery for all sampled planted partitions when ${\mu\leq 0.4}$ and for a subset of sampled planted partitions when ${\mu=0.5}$ and ${\mu=0.6}$. The results for multilayer \textsc{InfoMap} on temporal benchmarks with uniform interlayer dependencies with ${p=0.99}$ (see \cref{sfig:T_Inf_0.99} for ${\mu=0.3}$ and ${\mu=0.4}$) and $p=1$ (see \cref{sfig:T_Inf_1} for ${\mu=0.5}$ and ${\mu=0.6}$) are the only instances where we observe large differences in results for different partitions that are generated by our model with the same parameter values.

Comparing results for \textsc{GenLouvain} (see \cref{sfig:T_L_0.5,sfig:T_L_0.85,sfig:T_L_0.95,sfig:T_L_0.99,sfig:T_L_1,sfig:TCP_L_0.85,sfig:TCP_L_0.95,sfig:TCP_L_1}) and \textsc{GenLouvainRand} with reiteration and post-processing (see \cref{sfig:T_LRW_0.5,sfig:T_LRW_0.85,sfig:T_LRW_0.95,sfig:T_LRW_0.99,sfig:T_LRW_1,sfig:TCP_LRW_0.85,sfig:TCP_LRW_0.95,sfig:TCP_LRW_1}), we see that the difference in results between the two optimization heuristics for multilayer modularity is even more pronounced in these temporal examples than in the multiplex examples in \cref{sec:multiplex_examples}. In particular, \textsc{GenLouvain} exhibits an abrupt change in behavior near ${\omega=1}$; this is related to the transition behavior that was described in~\cite{Bazzi2014}. As we explained in \cref{sec:multiplex_examples} (where we also observed this phenomenon), this transition occurs for the following reason: for values of $\omega$ that are above a certain threshold~\cite{Bazzi2014}, only interlayer merges occur in the first phase of the Louvain heuristic. This abrupt transition no longer occurs with the additional randomization in \textsc{GenLouvainRand}, which exhibits smooth behavior as a function of $\omega$. However, unlike in our multiplex experiments, we observe benefits from the behavior of \textsc{GenLouvain} in some situations, especially for networks with weak community structure (specifically, for ${\mu\geq0.8}$). This phenomenon first becomes noticeable for networks with ${p=0.95}$, and it becomes more pronounced for progressively larger values of $p$. Compare \cref{sfig:T_L_0.95,sfig:T_L_0.99,sfig:T_L_1} with \cref{sfig:T_LRW_0.95,sfig:T_LRW_0.99,sfig:T_LRW_1}; and compare \cref{sfig:TCP_L_0.95,sfig:TCP_L_1} with \cref{sfig:TCP_LRW_0.95,sfig:TCP_LRW_1}.  

Our results for temporal benchmark networks with change points (see \cref{fig:tempCP-bench}) are mostly similar to those for temporal benchmark networks with uniform dependencies when considering $\mean{\NMI}$ to compare planted and recovered partitions. Only the results for $p=1$ are noticeably different. (Compare \cref{sfig:TCP_L_1,sfig:TCP_LRW_1,sfig:TCP_Inf_1} with \cref{sfig:T_L_1,sfig:T_LRW_1,sfig:T_Inf_1}.) Because of the change points, the induced planted partitions for each layer are not identical, even when $p=1$. Consequently, unlike in \cref{fig:temp-bench}, the results for $p=1$ in \cref{fig:tempCP-bench} are qualitatively similar to those for smaller values of $p$. In principle, one might expect multilayer \textsc{InfoMap} to have an advantage over multilayer-modularity-based methods in this experiment, as the former's quality function is designed to detect abrupt changes in community structure, whereas maximizing multilayer modularity always favors some persistence of community labels from one layer to the next. However, this theoretical advantage is not borne out in our experiment. When comparing the planted and recovered partitions using the NMI between multilayer partitions, we see that multilayer \textsc{InfoMap} can correctly recover the change points only for ${p=1}$ and $\mu=\{0,0.1\}$ (see \cref{app:mNMI}).

%%%%%%%%%%%%%%%%%%%%%%%%%%%%%%%%
%%%%%%%%%%%%%%%%%%%%%%%%%%%%%%%%

\subsection{Multi-aspect examples}
\label{sec:multiaspect_examples}

\begin{figure}
\subfloat[$\hat{p}=0.85$, $a=0.1$]{
\includegraphics[]{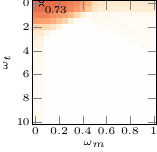}
}
\subfloat[$\hat{p}=0.85$, $a=0.5$]{
\includegraphics[]{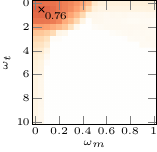}
}
\subfloat[$\hat{p}=0.85$, $a=0.9$]{
\includegraphics[]{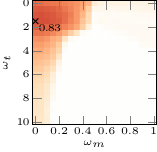}
}

\subfloat[$\hat{p}=0.95$, $a=0.1$]{
\includegraphics[]{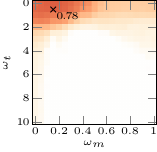}
}
\subfloat[$\hat{p}=0.95$, $a=0.5$]{
\includegraphics[]{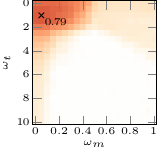}
}
\subfloat[$\hat{p}=0.95$, $a=0.9$]{
\includegraphics[]{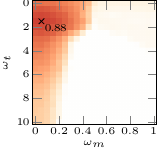}
}

\subfloat[$\hat{p}=0.99$, $a=0.1$]{
\includegraphics[]{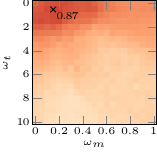}
}
\subfloat[$\hat{p}=0.99$, $a=0.5$]{
\includegraphics[]{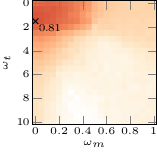}
}
\subfloat[$\hat{p}=0.99$, $a=0.9$]{
\includegraphics[]{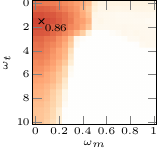}
}

\caption{Multi-aspect networks with uniform interlayer dependencies.
Effect of the strength of temporal interlayer coupling $\omega_t$ and multiplex interlayer coupling $\omega_m$ on the ability of maximization of multilayer modularity to recover the planted partition in two-aspect multilayer networks that we generate using our framework. 
The first aspect is temporal (and thus ordered) and the second aspect is multiplex (and thus unordered). 
The heat maps show the mean ($\mean{\mathrm{NMI}},\ 0\,\protect\includegraphics{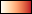}\,1$) over all layers of the NMI between induced planted partitions and recovered community structure. A black cross indicates the maximum value of $\mean{\mathrm{NMI}}$. The dependence between structure in different layers increases from top to bottom, and the importance of the temporal aspect versus the multiplex aspect increases from left to right. We average our results for each pair of parameters over $10$ runs of \textsc{GenLouvainRand} with reiteration. The generated networks have $n=100$ nodes and $l=100$ layers with $\vert\set{L}_\mathrm{T}\vert = \vert\set{L}_\mathrm{M}\vert=10$, so there are a total of $10000$ state nodes. We fix the community-mixing parameter of the M-DCSBM benchmark to be $\mu=0.5$.
}
\label{fig:multiaspect_example}
\end{figure}

In this section, we illustrate the ability of our framework to generate multilayer networks with more than one aspect. A common situation that gives rise to multilayer networks with two aspects is a multiplex network that changes over time (e.g., citations patterns that change over time~\cite{Starnini2017}).
In our framework, such a situation corresponds to the interlayer-dependency tensor in \cref{fig:coupling_temp_mult}. For the illustrative examples in this section, we use uniform multiplex and temporal dependencies; and we parametrize the interlayer dependency tensor as follows:
\begin{equation*}
	\begin{multlined}
\LP_{\alpha_t,\alpha_m}^{\beta_t,\beta_m}=\delta(\alpha_t,\beta_t)(1-\delta(\alpha_m,\beta_m))(1-a)\hat{p}/(l_m - 1) \\+ \delta(\alpha_t+1,\beta_t)\delta(\alpha_m,\beta_m) a \hat{p}\,,
	\end{multlined}
\end{equation*}
where $a$ controls the balance between multiplex and temporal dependencies and $\hat{p}$ controls the overall strength of the interlayer dependencies.

To recover communities, we optimize a multi-aspect generalization, 
\begin{equation}\label{eq:ma_modularity}
\begin{aligned}
	Q(\mat{S}) = \smash{\sum_{\substack{(i,\balpha) \in \set{V}_M \\ (j,\bbeta) \in \set{V}_M}}} \biggl[\Bigl(A^{j,\bbeta}_{i,\balpha} - \tfrac{k_{i,\balpha}^{\balpha}k_{j,\bbeta}^{\bbeta}}{2m_\balpha^\bbeta}\Bigr)&\delta(\balpha,\bbeta)\\
 +\ C_{\balpha}^{\bbeta}\,\delta(i,j)\,\biggr]&\delta(S_{i,\balpha},S_{j,\bbeta})\,,
\end{aligned}
\end{equation}
of the multilayer modularity function that was introduced in \cite{Mucha2010}.
 
We specify the \define{interlayer-coupling tensor} $\mat{C}$ (a multi-aspect generalization of the \quoting{interslice coupling} of~\cite{Mucha2010}), such that it reflects the planted interlayer dependencies, by writing
\begin{equation*}
	\begin{multlined}
C_{\alpha_t,\alpha_m}^{\beta_t,\beta_m} = \delta(\alpha_t,\beta_t)(1-\delta(\alpha_m,\beta_m))\omega_m\\+ \delta(\alpha_t+1,\beta_t)\delta(\alpha_m,\beta_m) \omega_t\;,
	\end{multlined}
\end{equation*}
{where $\omega_t$ denotes the coupling parameter for the temporal dependencies and $\omega_m$ denotes the coupling parameter for the multiplex dependencies}. 

In \cref{fig:multiaspect_example}, we illustrate the effects of the two interlayer coupling parameters, $\omega_t$ and $\omega_m$, on the extent to which multilayer modularity maximization can recover a planted partition. We use \textsc{GenLouvainRand} with reiteration but without post-processing to find a local optimum of \cref{eq:ma_modularity}. For all of the values of $\hat{p}$ and $a$ that we consider in \cref{fig:multiaspect_example}, we see some improvement in the performance of multilayer modularity maximization with nonzero interlayer coupling compared with the case where both $\omega_t=0$ and $\omega_m=0$. As the latter case corresponds to independently maximizing modularity on each layer of a network, this demonstrates that multilayer modularity is able to exploit the interlayer dependencies to better recover the planted partition. As expected, increasing $a$ {(which gives more weight to temporal dependencies)} leads to a shift of the region of the parameter space with good planted-partition recovery to larger values of $\omega_t$ and smaller values of $\omega_m$. For progressively larger $\hat{p}$, we observe a small overall increase in the value of $\mean{\text{NMI}}$, but its dependence on $\omega_t$ and $\omega_m$ remains similar.

%%%%%%%%%%%%%%%%%%%%%%%%%%%%%%%%
%%%%%%%%%%%%%%%%%%%%%%%%%%%%%%%%

\section{Conclusions}
\label{sec:conclusion}

We introduced a unifying and flexible framework for the construction of generative models for mesoscale structures in multilayer networks. The three most important features of our framework are the following: (1) it includes an explicitly parametrizable tensor $\mat{P}$ that controls interlayer-dependency structure; (2) it can generate an extremely general, diverse set of multilayer networks (including, e.g., temporal, multiplex, \quoting{multilevel}~\cite{Lomi2016}, and multi-aspect multilayer networks with uniform or nonuniform dependencies between state nodes); and (3) it is modular, as the process of generating a partition is separate from the process of generating edges, enabling a user to first generate a partition and then use any planted-partition network model. Along with our paper, we provide publicly available code~\cite{bazzi2019} that users can modify to readily incorporate different types of null distributions (see \cref{nulldistribution}), interlayer-dependency structures (see \cref{generalcase}), and planted-partition network models (see \cref{sec:sampling_networks}).

The ability to explicitly specify interlayer-dependency structure makes it possible for a user to control which layers depend directly on each other (by deciding which entries in the interlayer-dependency tensor are nonzero) and the strengths of such dependencies (by varying the magnitude of entries in the interlayer-dependency tensor). One can thereby generate multilayer networks with either a single aspect or multiple aspects (e.g., temporal and/or multiplex networks) and vary dependencies between layers from the extreme case in which induced partitions in a planted multilayer partition are the same across layers to the opposite extreme, in which induced partitions in a planted multilayer partition are generated independently for each layer from a null distribution for that layer. To the best of our knowledge, this level of generality is absent from existing generative models for mesoscale structures in multilayer networks, as those models tend to only consider networks with a single aspect (e.g., temporal or multiplex) or limited interlayer-dependency structures (e.g., with the same induced partitions in a planted multilayer partition across all layers). 

We illustrated several examples of generative models that one can construct from our framework, with a focus on a few special cases of interest, rather than on trying to discuss as many situations as possible.  We focused on community structure in our numerical experiments, because it is a commonly studied mesoscale structure, but one can also use our framework to construct generative models of intra-layer mesoscale structures other than community structure (e.g., core--periphery structure, bipartite structure, and so on) by taking advantage of the flexibility of our framework's ability to use any single-layer planted-partition network model. For our single-aspect examples, we assumed that interlayer dependencies exist either between all contiguous layers (a special case of temporal networks), between all layers (a special case of multiplex networks), or between all contiguous groups of layers (another special case of multiplex networks). For both our temporal and multiplex examples, we considered both uniform and nonuniform interlayer dependencies. We also combined some of these scenarios in an example with two aspects, a temporal aspect and a multiplex aspect. However, our framework's flexibility allows us to construct generative models of multilayer networks with more realistic features. For example, for temporal networks, one can introduce dependencies between a layer and all layers that follow it (such that \cref{sfig:temporal} is an upper triangular matrix with nonzero entries above the diagonal) to incorporate memory effects \cite{Rosvall2014}. One can also consider interlayer dependencies that are not layer-coupled. For example, dependencies can be diagonal but nonuniform, or they can be nonuniform and exist only between sets of related nodes.

In \cref{sec:numerical_examples}, we consider the commonly studied case of a multilayer network with only intralayer edges and connectivity patterns in the different layers that depend on each other. We use a slight variant of the DCSBM benchmark from~\cite{Karrer2011} to generate edges for each layer (see \cref{alg:dcsbm}). However, other types of multilayer networks are also important \cite{Kivela2014}, and one can readily combine our approach for generating multilayer partitions with different network generative models that capture various important features. For example, one can use an SBM to generate interlayer edges (e.g., using our M-DCSBM framework, which we discuss in \cref{app:MDCSBM}), or one can replace the degree-corrected SBM in \cref{sec:sampling_networks} with any other planted-partition network model or other interesting models (e.g., other variants of SBMs~\cite{Holland1983, Jacobs2014} and models for networks whose structure is affected by space (and perhaps spatially embedded) \cite{Sarzynska2014} or arbitrary latent features \cite{Newman2015}). In all of these examples, dependencies between connectivity patterns in different layers arise only from a planted multilayer partition. It is also possible to modify our network generation process (see \cref{sec:sampling_networks}) to incorporate additional dependencies between layers beyond those that are induced by a planted mesoscale structure, such as by introducing dependencies between a node's degree in different layers~\cite{Nicosia2015,Lee2012} or burstiness \cite{Lambiotte2013} in inter-event-time distributions of edges.

Our work has the potential for many useful and interesting extensions, and we highlight three of these. First, although we have given some illustrative numerical examples in \cref{sec:numerical_examples}, the area of benchmarking community-detection methods in multilayer networks is far from fully developed. Generative models are useful tools for understanding the behavior of community-detection methods in detail and thus for suggesting ways of improving heuristic algorithms without losing scalability. One can use our framework to construct benchmark models that provide a test bed for gaining insight into the advantages and shortcomings of community-detection methods and algorithms (and, more generally, of \mesostructure{}-detection methods and algorithms). Importantly, we expect these benchmark models to be very informative for detecting potential artifacts of algorithms that can sometimes be masked in real-world applications. Second, a well-understood generative model can be a powerful tool for statistical inference (i.e., inferring the structure of a multilayer network rather than generating a multilayer network with a planted structure)~\cite{Fortunato2016}. For temporal networks, closed forms for the joint distribution of \mesostructure{} assignments have been derived for models that are special cases of our framework~\cite{Ghasemian2015, Pamfil2018}. These results may be useful for statistical inference.
Additionally, it seems likely that it is possible
to adapt Bayesian inference techniques (including Gibbs sampling~\cite{Snijders1997, Nowicki2011, Peixoto2017b} or variational methods~\cite{Latouche2012}) that have been developed for SBMs both for inferring a multilayer partition and for inferring an interlayer-dependency tensor. A key advantage of the generality of the framework that we have developed in the present paper is that it may be possible to frame many model-selection questions in terms of posterior estimation of the interlayer-dependency tensor.
Finally, it is important to model interdependent data streams and not just fixed data sets. For example, for any fully-ordered multilayer network, our generative model respects the causality of layers (e.g., temporal causality), and one can thus update a multilayer network with a new layer without the need to update any previous layers. It is critical to develop generative models that are readily adaptable to such situations, and our work in this paper is a step in this direction.

%%%%%%%%%%%%%%%%%%%%%%%%%%%%%%%%
%%%%%%%%%%%%%%%%%%%%%%%%%%%%%%%%

\appendix

%%%%%%%%%%%%%%%%%%%%%%%%%%%%%%%%
%%%%%%%%%%%%%%%%%%%%%%%%%%%%%%%%

\section{Example null distributions}
\label{app:nulldistribution}

In this appendix, we discuss parameter choices for a categorical null distribution and give concrete examples that can be useful for modeling mesoscale structure in temporal or multiplex networks.

In \cref{nulldistribution}, we described the general form of a categorical null distribution:
\begin{equation}
	\PP_0^\balpha[s]=\begin{cases} p_s^{\balpha}\, ,& s\in\{1, \ldots, n_{\mathrm{set}}\}\,,\\
	0\, , & \text{otherwise\,,}\end{cases} 
	\label{null2}
\end{equation}
where $p_s^{\balpha}$ is the probability for a state node in layer $\balpha$ to be assigned to a \mesostructure{} $s$ in the absence of interlayer dependencies, $n_{\mathrm{set}}$ is the number of \mesostructure{}s in the multilayer partition, and $\sum_{s=1}^{n_{\mathrm{set}}} p_s^{\balpha}= 1$ for each $\balpha \in L$.  The support $\set{G}^{\balpha}$ of the null distribution $\PP_0^{\balpha}$ is  $\set{G}^{\balpha} = \{s: \PP_0^{\balpha}[s] \neq 0\}$. We say that a label $s$ is \define{active} in a layer $\balpha$ if it is in the support of the null distribution $\PP_0^\balpha$ (i.e., if $\PP_0^{\balpha}[s] \neq 0$); and we say that a label is \define{inactive} in layer $\balpha$ if it is in the complement of the support of $\PP_0^\balpha$ (i.e., if $\PP_0^{\balpha}[s] = 0$). 

A natural choice is to sample $\vec{p}^{\balpha}$ from a Dirichlet distribution, which is the conjugate prior for the categorical distribution \cite{Raiffa2000,Agarwal2010}. The Dirichlet distribution over $q$ variables has $q$ parameters $\theta_1, \ldots, \theta_q$. Its probability density function is 
\begin{equation*}
	p(x_1, \ldots , x_q) = \frac{\gamdist \left(\sum_{i=1}^q\theta_i\right)}{\prod_{i=1}^q \gamdist(\theta_i)} \prod_{i=1}^q x_i^{\theta_i-1}\, ,
\end{equation*} 
where $x_i \in (0,1)$ and $\theta_i > 0$ for each $i\in\{1,\ldots, q\}$. The case in which all $\theta_i$ are equal is called a \quoting{symmetric Dirichlet distribution}, which we parametrize by the common value $\theta$ (the so-called \quoting{concentration parameter}) of the parameters and the number $q$ of variables. 

The concentration parameter $\theta$ determines the types of discrete probability distributions that one is likely to obtain from the symmetric Dirichlet distribution. For $\theta=1$, the symmetric Dirichlet distribution is the continuous uniform distribution over the space of all discrete probability distributions with $n_{\mathrm{set}}$ states. As $\theta \to \infty$, the Dirichlet distribution becomes increasingly concentrated near the discrete uniform distribution, such that all entries in $\vec{p}^\balpha$ are approximately equal. As $\theta \to 0$, it becomes increasingly concentrated away from the uniform distribution, such that $\vec{p}^\balpha$ tends to have $1$ (or a few) large entries and all other entries are close to $0$. Consequently, to have very heterogeneous \mesostructure{} sizes, one would choose $\theta \approx 1$. To have all \mesostructure{}s be of similar sizes, one would choose a large value of  $\theta$. To have a few large \mesostructure{}s and many small \mesostructure{}s, one would choose $\theta$ to be sufficiently smaller than $1$. The value of $n_{\mathrm{set}}$ also affects the amount of \mesostructure{} label overlap across layers in the absence of interlayer dependencies. For example, if $\vec{p}^\balpha$ is the same for all layers, then larger values of $n_{\mathrm{set}}$ incentivize less label overlap across layers (because there are more possible labels for each layer) and smaller values of $n_{\mathrm{set}}$ incentivize more label overlap across layers (because there are fewer possible labels for each layer). 

In some situations --- e.g., when modeling the birth and death of communities in temporal networks or the appearance and disappearance of communities in multiplex networks --- it is desirable to have \mesostructure{} labels that have a nonzero probability in~\eqref{null2} only in some layers. For these situations, we suggest sampling the support of the distributions before sampling the probabilities $\vec{p}^{\balpha}$.
Given the supports for each layer, one then samples the corresponding probabilities from a symmetric Dirichlet distribution (or any other distribution over categorical distributions). That is,
\begin{equation}
	\vec{p}^\balpha_{\set{G}^{\balpha}} \sim \operatorname{Dir}(\theta, \left|\set{G}^{\balpha}\right|)\, , \qquad \vec{p}^\balpha_{\comp{G}^\balpha} = 0\, ,
\end{equation}
where $\comp{\set{G}}^{\balpha}=\{ s \in \{1, \ldots, n_{\mathrm{set}}\}: \PP_0^{\balpha}[s]=0\}$ is the complement of the support, with $n_{\mathrm{set}}=\max_{\balpha \in L} \max(\set{G}^{\balpha})$.

A simple example for a birth/death process for \mesostructure{}s is the following. First, fix a number of \mesostructure{}s and a support (i.e., active community labels) for the first layer. One then sequentially initializes the supports for the other layers by removing each \mesostructure{} 
in the support of the previous layer with probability $r_d\in [0,1]$ and adding a number, sampled from a Poisson distribution with rate $r_b\in[0,\infty)$, of new \mesostructure{}s (with new labels that are not active in any previous layer). In temporal networks, for example, this ensures that if a \mesostructure{} label is not in the support of a given layer, then the label is also not in the support of any subsequent layers. For this process, the expected number $\mean{|\set{G}^{\balpha}|}$ of \mesostructure{}s in a layer approaches $r_b/r_d$ as one iterates through the layers.
Therefore, one should initialize the size of the support for the first layer close to this value to avoid transients in the number of \mesostructure{}s. For this process, the lifetime of \mesostructure{}s follows a geometric distribution. The nature of the copying process that we use to introduce dependencies between induced partitions in different layers typically implies that \mesostructure{}s that have been removed from the support of the null distribution do not lose all of their members instantly, but instead shrink at a speed that depends on the values of the copying probabilities in the interlayer-dependency tensor.

One can also allow labels to appear and disappear when examining multiplex partitions. For example, given a value for $n_{\mathrm{set}}$, one can generate the support for each layer by allowing each label $s\in\{1,\ldots,n_{\mathrm{set}}\}$ to be present with some probability $\tilde{q}$ and absent with complementary probability $1-\tilde{q}$. This yields a sets of active and inactive \mesostructure{} labels for each layer. One can then sample the nonzero probabilities in $\vec{p}^\balpha$ that correspond to active labels from a Dirichlet distribution and set $p^\balpha_s$ to $0$ for each inactive label $s$. Because multiplex partitions are unordered, there is no notion of one layer occurring after another one, so we do not need to ensure that an inactive label in a given layer is also inactive in ``subsequent'' layers.

%%%%%%%%%%%%%%%%%%%%%%%%%%%%%%%%
%%%%%%%%%%%%%%%%%%%%%%%%%%%%%%%%

\section{Partition sampling process}
\label{app:sampling}

In this appendix, we provide a detailed description of the way in which we sample multilayer partitions, including a discussion of the convergence properties of our sampling process. As we mentioned in \cref{generalcase}, we assume that the multilayer partitions are generated by a copying process on the \mesostructure{} assignment of state nodes. Recall that the conditional probability distribution for the updated \mesostructure{} assignment of a state node, given the current state of the copying process (\cref{eq:updating} in \cref{generalcase}), is 
\begin{equation}\label{eq:updating2}
\begin{aligned}
	&\PP[S_{j,\bbeta}(\tau+1) = s|\mat S(\tau)]\\[5pt]
	&\quad \qquad =\smashoperator{\sum_{{(i,\balpha)\in \set{V}_M}}}P_{i,\balpha}^{j,\bbeta}\,\delta(S_{i,\balpha}(\tau),s)\\[5pt]
	&\quad\qquad \qquad + \left({1-\hat{p}_{j,\bbeta}}\right)\PP_0^{\bbeta}[S_{j,\bbeta} = s]\,,\\
	&\hat{p}_{j,\bbeta} = \smashoperator{\sum_{(i,\balpha)\in \set{V}_M}} P_{i,\balpha}^{j,\bbeta}\, ,
	\end{aligned} 
\end{equation}
where $\PP_0$ is a given set of independent layer-specific null distributions and $\mat{P}$ is a given interlayer-dependency tensor. Further recall that a multilayer network can have both ordered aspects and unordered aspects, where we denote the set of ordered aspects by $\set{O}$ and the set of  unordered aspects by $\set{U}$. Our goal is to sample multilayer partitions that are consistent with generation
by \cref{eq:updating2} while respecting the update order from its
ordered aspects (if present). Recall that two layers, $\balpha$ and $\bbeta$, are order-equivalent (which we denote by ${\balpha\sim_{\set{O}} \bbeta}$) if ${\alpha_a=\beta_a}$ for all ${a \in \set{O}}$. Based on this equivalence relation, we obtain an ordered set of equivalence classes $\set{L}/{\sim_{\set{O}}}$ by inheriting the order from the ordered aspects (where, for definiteness, we consider the lexicographic order~\cite{note_lexi} over ordered aspects). We want to simultaneously sample the \mesostructure{} assignment of state nodes within each class of order-equivalent layers and we want to sequentially sample the \mesostructure{} assignment of state nodes in non-order-equivalent layers, conditional on the fixed assignment of state nodes in preceding layers (see \cref{alg:sample_partition}).

\begin{algorithm}
\begin{algorithmic}
\raggedright
\Function{SamplePartition}{$\mat{P}$, $\PP_0$, $\set{O}$, $n_{\mathrm{upd}}$}
\For{$(i, \balpha) \in V_m$}
\State $S_{i,\balpha}\sim\mathbb{P}_0^{\balpha}$\Comment{Initialize partition tensor using null distributions}
\EndFor
\For{$[\balpha] \in \set{L}/{\sim_\set{O}}$} \Comment{Loop over ordered aspects in lexicographic order}
\For{$\tau \in \{1,\ldots |[\balpha]|\,n_{\mathrm{upd}}\}$}
\State $\balpha\sim\mathop{U}([\balpha])$ \Comment{Sample uniformly from order-equivalent layers}
\For{$i \in V$} \Comment{Loop over nodes in some order}
\State $S_{i,\balpha} \sim \PP[S_{i,\balpha}|\mat{S}]$ \Comment{Update $(i,\balpha)$ according to \cref{eq:updating2}}
\EndFor
\EndFor
\EndFor

\State \Return{$\mat{S}$}
\EndFunction
\end{algorithmic}

\caption{Pseudocode for generating multilayer partitions with interlayer-dependency tensor $\mat{P}$, null distributions $\PP_0^{\balpha}$, and ordered aspects $\set{O}$. The parameter $n_{\mathrm{upd}}$ is the expected number of updates for each state node.
\label{alg:sample_partition}
}
\end{algorithm}

%%%%

\subsection*{Scan order and compatibility}

To sample the induced partitions for each class of order-equivalent layers, we use a technique that resembles Gibbs sampling \cite{Gelfand2000}. As in Gibbs sampling, we define a Markov chain on the space of multilayer partitions by updating the \mesostructure{} assignment of state nodes by sampling from \cref{eq:updating2}. In Gibbs sampling, one assumes that the conditional distributions that are used to define this Markov chain are compatible, in the sense that there exists a joint distribution that realizes these conditional distributions~\cite{Hobert1998, Arnold1989, Chen2010, Chen2014}. However, for a general interlayer-dependency tensor, there is no guarantee that the distributions that are defined by \cref{eq:updating2} are compatible. Following~\cite{Heckerman2000}, we use the term ``pseudo-Gibbs sampling'' to refer to Gibbs sampling from a set of potentially incompatible conditional distributions. We can define different Markov chains by changing the way in which we select which conditional distribution to apply at each step (i.e., which state node to update at each step). The order in which one cycles through the conditional distributions is known as the \define{scan order}~\cite{Kuo2017}. We use the term \define{sampling Markov chain} for the Markov chain that is defined by pseudo-Gibbs sampling with a specified scan order. Common scan orders are cycling through conditional distributions in a fixed order and sampling the update distribution uniformly at random from the set of conditional distributions. In pseudo-Gibbs sampling, the choice of scan order deserves careful attention. In pseudo-Gibbs sampling, different {scan orders} correspond to sampling from different distributions~\cite{Kuo2017}. (By contrast, in Gibbs sampling, the choice of {scan order} influences only the speed of convergence.) In particular, when cycling through conditional distributions in a fixed order, conditional distributions that are applied later have an outsized influence on the stationary distributions of the sampling Markov chain~\cite{Kuo2017}. Sampling the update distribution uniformly at random mitigates this problem, at the expense of introducing computational overhead. 

We can improve on a purely random sampling strategy by exploiting the structure of the interlayer-dependency tensor. In particular, note that the conditional distributions for state nodes in the same layer are independent (as we assume that the interlayer-dependency tensor has no intralayer contributions) and hence compatible. When updating a set of state nodes from a given layer, we can update them in any order or even update them concurrently. To take advantage of this fact, we first sample a layer uniformly at random from the current class of order-equivalent layers and then update all state nodes in that layer. The sequence of layer updates defines a Markov chain on the space of multilayer partitions, and its convergence properties determine the level of effectiveness of our sampling algorithm. 

%%%%%%%%%%%%%%%%%%%%%%%%%%%%%%%%
%%%%%%%%%%%%%%%%%%%%%%%%%%%%%%%%

\subsection*{Convergence guarantees}

We now discuss two key properties --- aperiodicity and ergodicity --- that determine the convergence behavior of finite Markov chains. 

First, note that sampling layers uniformly at random guarantees that the Markov chain is aperiodic. To see this, note that a sufficient condition for aperiodicity is that, for all states, there is a nonzero probability to transition from the state to itself. 
This clearly holds for our sampling chain, as it is possible for two consecutive updates to update the same layer, and (because the probability distributions in \cref{eq:updating2} remain unchanged at the second update) there is a positive probability that the second update does not change the partition. 

Second, note that if ${\hat{p}_{i,\balpha} < 1}$ for all state nodes ${(i, \balpha)\in \set{V}_M}$ such that ${\balpha \in [\balpha]}$, where $[\balpha]$ is a class of order-equivalent layers, then the sampling Markov chain for that class is ergodic over a subspace of multilayer partitions that includes the support ${\set{G}^{[\balpha]} = \prod_{\balpha \in [\balpha]} \set{G}^{\balpha}}$ of the null distributions. We have already shown that the sampling Markov chain is aperiodic, so all that remains is to verify that there exists a sample path from any arbitrary initial partition $\set{S}^0$ to any arbitrary multilayer partition ${\set{S}^g \in \set{G}^{[\balpha]}}$ in the support of the null distributions. Such a sample path clearly exists, as 
\begin{equation*}
	\begin{multlined}
		\PP\left[{S_{i,\balpha}(\tau+1) = S^g_{i,\balpha}} \;|\; {S_{i, \balpha}(\tau)=S^0_{i,\balpha}}\right] \\
\qquad \geq \frac{1}{|[\balpha]|} (1-\hat{p}_{i,\balpha}) \PP_0^{\balpha}[S^g_{i,\balpha}]>0 \,.
	\end{multlined}
\end{equation*}
Therefore, one can achieve the desired transition using one update for each state node.

Sample paths of a finite, aperiodic Markov chain converge in distribution to a stationary distribution of the Markov chain. However, different sample paths can converge to different distributions, and the eventual stationary distribution can depend both on the initial condition and on the transient behavior of the sample path. If the Markov chain is ergodic, the stationary distribution is unique, so it is independent of the initial condition and transient behavior.

After sufficiently many updates (i.e., by specifying a sufficiently large value for the expected number $n_{\mathrm{upd}}$ of updates for each state node in \cref{alg:sample_partition}), we approximately sample the state of the Markov chains for each class of order-equivalent layers from their stationary distributions. For fully-ordered multilayer networks (i.e., for multilayer networks with $\set{U}=\emptyset$), setting ${n_{\mathrm{upd}}=1}$ is sufficient for convergence, because the equivalence classes include only a single layer in this case and subsequent updates are thus independent samples from the same distribution.

Sampling initial conditions in an \paraphrase{unbiased} way can be important for ensuring that one successfully explores the space of possible partitions, as the updating process is not necessarily ergodic over the support of the null distributions when $\sum_{(i,\balpha)\in \set{V}_M}P_{i,\balpha}^{j,\bbeta}=1$ for some state nodes. For example, if $\hat{p}_{i,\balpha}=1$ for all $(i,\balpha)\in \set{V}_M$, then any partition in which all state nodes in each weakly connected component of the interlayer-dependency network have the same \mesostructure{} assignment is an absorbing state of the Markov chain.

However, provided the updating process has converged, any partition that one generates 
should still reflect the desired dependencies between induced partitions for different layers.
Therefore, to circumvent the problem of non-unique stationary distributions when the updating process is not ergodic, we reinitialize the updating process by sampling from the null distribution for each partition that we want to generate. Reinitializing the initial partition from a fixed distribution for each sample always defines a unique joint distribution, which is a mixture of all possible stationary distributions when the updating process is not ergodic. When the updating process is ergodic, it is equivalent to sample partitions by reinitializing or to sample multiple partitions from a single long chain. Using a single long chain is usually more efficient when one needs many samples and it is not problematic if there is some dependence between samples~\cite{Tierney1994}. 
In our case, however, we usually need only a few samples with the same parameters, and ensuring independence tends to be more important than ensuring perfect convergence (provided the generated partitions exhibit the desired interlayer-dependency structure). Using multiple chains thus has clear advantages even when the updating process is ergodic, and it is necessary to use multiple chains when it is not ergodic.

%%%%%

\subsection*{Convergence tests}

\begin{figure}
\centering
\includegraphics{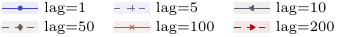}\\
\subfloat[$\hat{p}=0.5$]{
\includegraphics{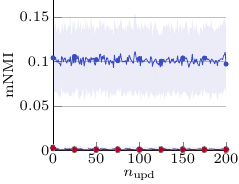}%
}%
\hfill
\subfloat[$\hat{p}=0.85$]{
\includegraphics{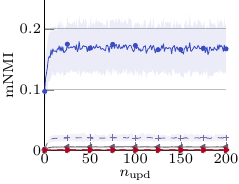}%
}\\

\subfloat[$\hat{p}=0.95$]{
\includegraphics{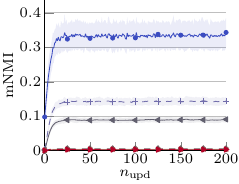}%
}%
\hfill
\subfloat[$\hat{p}=0.99$]{
\includegraphics{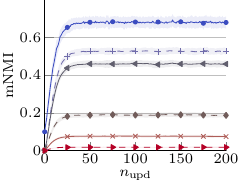}%
}\\
\subfloat[$\hat{p}=1$]{
\includegraphics{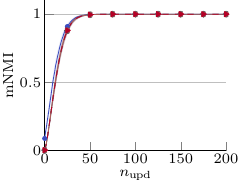}%
}%

\caption{Autocorrelation of the sampling Markov chain, as measured by mNMI. On the horizontal axis, we show the value of $n_{\mathrm{upd}}$ that we use to generate the first partition in the comparison. The lag is the number of additional updates per state node that we use to generate the second partition in the comparison. We average results over a sample of $100$ independent chains with different initial partitions. The shaded area indicates one standard deviation above and below the mean. Note that the large amount of noise for small values of the lag is not a result of differences between chains in the sample; instead, it is an inherent feature of all chains in the sample.\label{fig:conv-nmi}}
\end{figure}

\begin{figure}
\includegraphics{fig/convergence_0.5_nmi_l.pdf}\\

\subfloat[$\hat{p}=0.5$]{\includegraphics{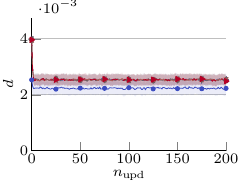}%
}%
\hfill
\subfloat[$\hat{p}=0.85$]{
\includegraphics{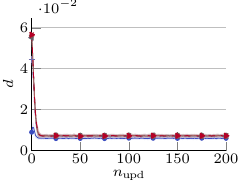}%
}\\

\subfloat[$\hat{p}=0.95$]{
\includegraphics{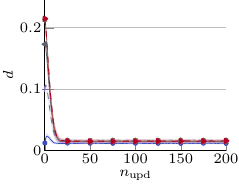}%
}%
\hfill
\subfloat[$\hat{p}=0.99$]{
\includegraphics{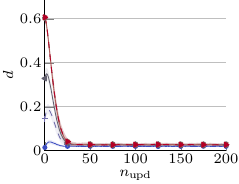}%
}\\

\subfloat[$\hat{p}=1$]{
\includegraphics{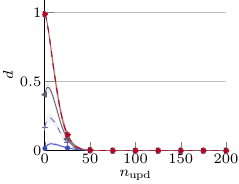}%
}

\caption{Convergence of the dependency pattern between induced partitions in different layers, as measured by the mean absolute distance $d$ between NMI matrices (see \cref{eq:d}). On the horizontal axis, we show the value of $n_{\mathrm{upd}}$ that we use to generate the first partition in the comparison. The lag  is the number of additional updates per state node that we use to generate the second partition in the comparison.  We average results over a sample of $100$ independent chains with different initial partitions. The shaded area indicates one standard deviation above and below the mean. \label{fig:conv-distance}}
\end{figure}

It is difficult to determine whether a Markov chain has converged to a stationary distribution, mostly because it is difficult to distinguish the case of a slow-mixing chain becoming stuck in a particular part of state space from the case in which the chain has converged to a stationary distribution. There has been much work on trying to define convergence criteria for Markov chains~\cite{Cowles1996}, but none of the available approaches are entirely successful. In practice, one usually runs a Markov chain (or chains) for a predetermined number of steps. One checks manually for a few examples that the resulting chains exhibit behavior that is consistent with convergence by examining autocorrelations between samples of the same chain and cross-correlations between samples of independent chains with the same initial state. When feasible, one can also check whether parts of different, independent chains or different parts of the same chain are consistent with sampling from the same distribution.

To estimate the number of updates that we need in the numerical experiments of \cref{sec:multiplex_examples}, we examine the autocorrelation of the Markov chain in two different ways. In \cref{fig:conv-nmi}, we consider the multilayer NMI (mNMI) between sampled partitions at different steps of the Markov chain. We are interested in the behavior of the mNMI both as a function of $n_{\mathrm{upd}}$ (the expected number of updates per state node that we use to generate the first partition in the comparison) and the lag (the number of additional updates per state node that we use to generate the second partition in the comparison). As the Markov chain converges to a stationary distribution, the value of the mNMI for a given lag should become independent of $n_{\mathrm{upd}}$ (so the curves in \cref{fig:conv-nmi} should become flat). The results in \cref{fig:conv-nmi} suggest that the number of updates that we need for convergence increases moderately with $\hat{p}$, where ${n_{\mathrm{upd}}\approx 50}$ is sufficient for convergence even when ${\hat{p}=1}$. In \cref{fig:conv-distance}, we examine whether the \define{dependency pattern} between induced partitions in different layers --- we characterize such a pattern by the NMI between induced partitions in different layers --- has converged using the mean absolute distance
\begin{equation}\label{eq:d}
	\begin{multlined}
d(\set{S},\set{T})=\\
\tfrac{2}{l(l-1)} \sum_{\alpha<\beta} \left|\operatorname{NMI}(\set{S}\vert_\alpha,\set{S}\vert_\beta)-\operatorname{NMI}(\set{T}\vert_\alpha,\set{T}\vert_\beta)\right|
\end{multlined}
\end{equation}
between the NMI matrices between induced partitions in a multilayer partition. (We showed examples of such NMI matrices in \cref{fig:temporalpartitions_ex,fig:multiplexpartitions_ex}.) As with
mNMI, the mean absolute distance $d$ should become independent of $n_{\mathrm{upd}}$ once the Markov chain has converged to a stationary distribution. This yields estimates for the minimum number of updates for convergence that are consistent with those from the results in \cref{fig:conv-nmi}. It is always safe to use a larger number of updates to sample partitions, and the results from \cref{fig:conv-distance,fig:conv-nmi} suggest that our choice of ${n_{\mathrm{upd}}=200}$ is conservative.

\Cref{fig:conv-nmi,fig:conv-distance} reveal additional important information about the behavior of our sampling process. We can estimate the mixing time of the sampling Markov chain at stationarity from \cref{fig:conv-nmi}, because the mixing time corresponds to the lag that is necessary for the mNMI to converge to $0$. The mixing time at stationarity increases with $\hat{p}$ and becomes infinite as ${\hat{p} \rightarrow1}$, because the Markov chain converges to an absorbing state. From \cref{fig:conv-distance}, we observe that, for all values of $\hat{p}$, the mean absolute distance $d$ between NMI matrices converges to a value near $0$ and becomes essentially independent of the lag. This corroborates our assumption that different partitions that we generate from our sampling process have similar interlayer-dependency structure. 

%%%%%%%%%%%%%%%%%%%%%%%%%%%%%%%%
%%%%%%%%%%%%%%%%%%%%%%%%%%%%%%%%

\section{A generative model for temporal networks}
\label{app:temporal}

In this appendix, we examine the generative model for temporal networks that we considered in \cref{specialcase}. In this model, we assume uniform dependencies between contiguous layers (i.e., $p_\beta = p \in[0,1]$ for all $\beta\in\{2,\ldots,l\}$ in \cref{sfig:temporal}). For this choice of interlayer-dependency tensor, our generative model reduces to  \cref{alg:temporalbenchmark}. In this case (and, more generally, for any fully-ordered multilayer network), convergence is automatic, because we need only one iteration through the layers. 

A first important feature of the generative model in \cref{alg:temporalbenchmark} is that it respects the arrow of time. In particular, \mesostructure{} assignments in a given layer depend only on \mesostructure{} assignments in the previous layer (e.g., the previous temporal snapshot) and on the null distributions $\mathbb{P}_0$. That is, for all $s\in\{1,\ldots,nl\}$ and all $\alpha\in\{2,\ldots,l\}$, we have
\begin{equation}
	\begin{aligned}
		\mathbb{P}\left[\left. S_{i,\alpha} = s \right|\,\set{S}\vert_{\alpha-1}\right] &= p\delta(S_{i,\alpha-1},s) \\
&\quad+ (1-p)\mathbb{P}_0^{\alpha}\left[S_{i,\alpha} = s\right]
\,,
	\end{aligned}
\label{eq:conditionalprob}
\end{equation}
where $\set S\vert_\alpha$ is the $n$-node partition that is induced on layer $\alpha$ by the multilayer partition $\set S$. The value of $p$ determines the relative importance of the previous layer and the null distribution. When $p=0$, \mesostructure{} assignments in a given layer depend only on the null distribution of that layer (i.e., on the second term of the right-hand side of \cref{eq:conditionalprob}). When $p=1$, \mesostructure{} assignments in a given layer are identical to those of the previous layer (and, by recursion, to \mesostructure{} assignments in all previous layers). 

Using \cref{eq:conditionalprob}, we see that the marginal probability that a given state node has a specific \mesostructure{} assignment is
\begin{align*}
	\mathbb{P}\left[S_{i,\alpha} = s\right]  &= p\mathbb{P}\left[S_{i,\alpha -1} = s\right]\\
	&\qquad+ (1-p)\mathbb{P}_0^{\alpha}\left[S_{i,\alpha} = s\right]\,,\\
	&= \mathbb{P}_0^1\left[S_{i,1} = s\right]p^{\alpha-1}\\
	 &\qquad+ (1-p)\sum_{\beta=2}^{\alpha-1} \mathbb{P}_0^\beta\left[S_{i,\beta} = s\right]p^{\alpha-\beta}\\ 
	 &\qquad+ (1-p)\mathbb{P}_0^\alpha\left[S_{i,\alpha} = s\right]\,,\quad \alpha > 1\,. \label{update}
\end{align*}
Computing marginal probabilities can be useful for computing expected \mesostructure{} sizes for a given choice of null distributions.

\begin{algorithm}
\begin{algorithmic}
\raggedright
\Function{TemporalPartition}{$p$, $\PP_0$}

\State $\mat{S}=$\textbf{ 0} \Comment{Initialize partition tensor of appropriate size} 

\For{$i \in V$} \Comment{Loop over nodes in some order}
\State $S_{i,1}\sim\mathbb{P}_0^1$ \Comment{Initialize induced partition on first layer using null distribution} 
\EndFor

\For{$\alpha \in 2\dots l$} \Comment{Loop over layers in sequential order}
\For{$i \in V$} \Comment{Loop over nodes in some order}

\If{$rand() < p $} \Comment{With probability $p$}
\State $S_{i,\alpha} = S_{i,\alpha-1}$ \Comment{Copying update (Step C)}
\Else \Comment{With probability $1-p$}
\State $S_{i,\alpha} \sim\mathbb{P}_0^{\alpha}$ \Comment{Reallocation update (Step R)}
\EndIf

\EndFor
\EndFor

\State \Return{$\mat{S}$}
\EndFunction
\end{algorithmic}

\caption{Pseudocode for generating partitions of a temporal network with uniform interlayer dependencies $p$ between successive layers (i.e., $p_\beta= p$ for all $\beta\in\{1,\ldots,l\}$ in  \cref{sfig:temporal}). 
\label{alg:temporalbenchmark}
}
\end{algorithm}

We now highlight how the copying update (i.e., Step C) and the reallocation update (i.e., Step R) in \cref{alg:temporalbenchmark} govern the evolution of \mesostructure{} assignments between consecutive layers. Steps C and R deal with the movement of nodes by first removing some nodes (``subtraction'') and then reallocating them (``addition''). In Step C, a \mesostructure{} assignment $s$ in layer $\alpha$ can lose a number of nodes that ranges from $0$ to all of them. It can keep all of its nodes in layer $\alpha+1$ (i.e., $\set{S}_s\vert_{\alpha + 1} = \set{S}_s\vert_{\alpha}$), lose some of its nodes (i.e., $\set{S}_s\vert_{\alpha + 1} \subset \set{S}_s\vert_{\alpha}$), or disappear entirely (i.e., $\set{S}_s\vert_{\alpha + 1} = \varnothing$ and $\set{S}_s\vert_{\alpha} \neq \varnothing$). The null distribution in Step R is responsible for a \mesostructure{} assignment $s$ gaining new nodes (i.e., $\set{S}_s\vert_{\alpha + 1} \not\subset \set{S}_s\vert_{\alpha}$) or for the appearance of a new \mesostructure{} label (i.e., $\set{S}_s\vert_{\alpha + 1} \neq \varnothing$ and $\set{S}_s\vert_{\alpha} = \varnothing$). When defining the null distributions $\mathbb{P}_0$, it is necessary to consider the interplay between Step C and Step R.  

To illustrate how the \mesostructure{}-assignment copying process and the null distribution in \cref{alg:temporalbenchmark} can interact with each other, we give the conditional probability that a label disappears in layer $\alpha$ and the conditional probability that a label appears in layer $\alpha$. For all $s\in\{1,\ldots,nl\}$ and all $\alpha\in\{2,\ldots,l\}$, the conditional probability that a label disappears in layer $\alpha$ is 
\begin{equation} \label{express}
\begin{aligned}
	&\mathbb{P}\left[\left.\set{S}_s\vert_{\alpha} = \varnothing \right| \,\set{S}\vert_{\alpha-1} \right]  \\
	&\qquad = \left[(1-p)\left(1-\mathbb{P}_0^{\alpha}[S_{i,{\alpha}} = s]\right)\right]^{\bigl| \set{S}_s\vert_{\alpha-1}\bigr|}\\
	 &\qquad \times\left[p + (1-p)\left(1-\mathbb{P}_0^{\alpha}[S_{i,{\alpha}} = s]\right)\right]^{n - \bigl| \set{S}_s\vert_{\alpha-1}\bigr|}\,.
	 \end{aligned}
\end{equation}

The expression \eqref{express} depends only on our copying process and simplifies to $(1-p)^{\bigl| \set{S}_s\vert_{\alpha-1}\bigr|}$ when $\mathbb{P}_0^{\alpha}[\set{S}_{i,\alpha} = s] = 0$ (i.e., when the probability of being assigned to label $s$ is $0$ using the null distribution of layer $\alpha$). Furthermore, for progressively larger $\mathbb{P}_0^{\alpha}[\set{S}_{i,\alpha} = s]$, the probability that a label disappears is progressively smaller.

For all $s\in\{1,\ldots,nl\}$ and all $\alpha\in\{2,\ldots,l\}$, the conditional probability that a label appears in layer $\alpha$ is 
\begin{equation} \label{express2}	
	\begin{aligned}
	&\mathbb{P}\big[\left.\set{S}_s\vert_{\alpha} \neq \varnothing \right| \,\set{S}_s\vert_{\alpha-1} = \varnothing \big]\\
&\quad = 1 - \bigg[p+(1-p)\bigg(\sum_{\set{S}_r\vert_{\alpha-1}\in \set{S} \vert_{\alpha-1}} \mathbb{P}_0^{\alpha} [S_{i,\alpha} = r]\bigg)\bigg]^{n}\,.
	\end{aligned}
\end{equation}
The expression \eqref{express2} gives the probability that at least one node in layer $\alpha$ has the label $s$, given that no node in layer $\alpha-1$ has the label $s$. When $\sum_{\set{S}_r\vert_{\alpha-1}\in \set{S}\vert_{\alpha-1}}\mathbb{P}_0^{\alpha}[S_{i,{\alpha}} = r] = 0$, the probability that a label appears depends only on our \mesostructure{}-assignment copying process and is given by $1 - p^n$. Furthermore, larger values of $\sum_{\set{S}_r\vert_{\alpha-1}\in \set{S}\vert_{\alpha-1}}\mathbb{P}_0^{\alpha}[S_{i,\alpha} = r]$ reduce the probability that a label appears in layer $\alpha$. 

All discussions thus far in this appendix hold for any choice of $\mathbb{P}_0$. In the next two paragraphs, we give two examples to illustrate some features of the categorical null distribution from \cref{nulldistribution}. In particular, we focus on the effect of the support of a categorical null distribution on a sampled multilayer partition. (See \cref{nulldistribution} for definitions of some of the notation that we use below.) The support $\set{G}^{\balpha}$ of a categorical null distribution $\PP_0^{\balpha}$ is  
$\set{G}^{\balpha} = \{s: p_s^{\balpha} \neq 0\}$, where $s \in\{1,\ldots,n_{\mathrm{set}}\}$. An important property of the support for \cref{sfig:temporal}
is that overlap between $\set{G}^{\alpha}$ and $\set{G}^{\alpha+1}$ (i.e., $\set{G}^{\alpha+1}\cap \set{G}^{\alpha}\neq \varnothing$) is a necessary condition for a \mesostructure{} in layer $\alpha$ to gain new members in layer $\alpha+1$.

Let $\vec{c}^{\alpha}$ denote the vector of expected induced \mesostructure{} sizes in layer $\alpha$ (i.e., $c^{\alpha}_s = n p^{\alpha}_s$), and suppose that the probabilities $\vec{p}^{\alpha}$ are the same in each layer (i.e., $\vec{p}^{\alpha} = \vec{p}$ for all $\alpha$). The expected number of \mesostructure{} labels is then the same for each layer, and the expected number of nodes with \mesostructure{} label $s$ is also the same in each layer and is given by $c^{\alpha}_s$. This choice produces a temporal network in which nodes change \mesostructure{} labels across layers in a way that preserves both the expected number of induced \mesostructure{}s in a layer and the expected sizes of induced \mesostructure{}s in a layer.

Now suppose that we choose the $\vec{p}^\alpha$ values such that their supports do not overlap (i.e., $\set{G}^{\alpha}\cap \set{G}^{\beta} = \varnothing$ for all $\alpha\neq \beta$). At each  Step C in \cref{alg:temporalbenchmark}, an existing \mesostructure{} label can then only lose members; and with probability $1-p^n$, at least one new label will appear in each subsequent layer. For this case, one expects $p c^{\alpha}_s$ members of \mesostructure{} $s$ in layer $\alpha$ to remain in \mesostructure{} $s$ in layer $\alpha+1$ and $(1-p) c^{\alpha}_s$ members of \mesostructure{} $s$ in layer $\alpha$ to be assigned to new \mesostructure{}s (because labels are nonoverlapping) in layer $\alpha+1$. This choice thus produces multilayer partitions in which the expected number of new \mesostructure{} labels per layer is nonzero (unless $p=1$) and the expected size of a given induced \mesostructure{} decreases in time.

%%%%%%%%%%%%%%%%%%%%%%%%%%%%%%%%
%%%%%%%%%%%%%%%%%%%%%%%%%%%%%%%%

\section{M-DCSBM sampling procedure}
\label{app:MDCSBM}

\begin{algorithm}
\begin{algorithmic}
\raggedright
\Function{DCSBM}{$\set{S},\vec{\sigma},\mat{W}$}
\State $\mat{A}=$\textbf{ 0}  \Comment{Initialize adjacency tensor of appropriate size} 
\For{$\balpha \in L$} \Comment{Loop over layers}
\For{$r \in 1 \ldots |\set{S}|$}
\For{$s \in r \ldots |\set{S}|$}
\State \Comment{Sample number of edges from a Poisson distribution}
\If{$r=s$}
\State $m=\Call{Poisson}{W_{r,\balpha}^{r,\balpha}/2}$
\Else
\State $m=\Call{Poisson}{W_{r,\balpha}^{s,\balpha}}$
\EndIf
\State $e=0$ \Comment{Count sampled edges}
\While{$e<m$}

\State 
\State $i=\Call{Sample}{\set{S}_r|_{\balpha},\vec{\sigma}_{\set{S}_r|_{\balpha},\balpha}}$\Comment{Sample node $i$ from induced community $r$ in layer $\balpha$ with probability $\sigma_{i,\balpha}$}
\State $j=\Call{Sample}{\set{S}_s|_{\balpha},\vec{\sigma}_{\set{S}_s|_{\balpha},\balpha}}$\Comment{Sample node $j$ from induced community $s$ in layer $\balpha$ with probability $\sigma_{j,\balpha}$}

\If{$i \neq j \And A_{i,\balpha}^{j,\balpha}=0$} \Comment{Reject self-edges or multi-edges}
\State $A_{i,\balpha}^{j,\balpha} = 1,\; A_{j,\balpha}^{i,\balpha} = 1$
\State $e=e+1$
\EndIf
\EndWhile
\EndFor
\EndFor
\EndFor
\State \Return{$\mat{A}$}
\EndFunction
\end{algorithmic}
\caption{Sampling multilayer networks from a DCSBM with community assignments $\set{S}$, node parameters $\vec{\sigma}$, and block tensor $\mat{W}$. 
\label{alg:dcsbm}}
\end{algorithm}

In this appendix, we discuss a  generalization to multilayer networks of the degree-corrected SBM (DCSBM) from Ref.~\cite{Karrer2011}. In other words, it is a multilayer DCSBM (M-DCSBM). The parameters of a general, directed M-DCSBM are a multilayer partition $\set{S}$ (which determines the assignment of state nodes to \mesostructure{}s), a block tensor $\mat{W}$ (which determines the expected number of edges between \mesostructure{}s in different layers), and a set $\mat{\sigma}$ of state-node parameters (which determine the allocation of edges to state nodes in \mesostructure{}s).
The probability of observing an edge (or the expected number of edges, if we allow multi-edges) from state node $(i,\balpha)$ to state node $(j,\bbeta)$ with \mesostructure{} assignments $r=S_{i,\balpha}$ and $s=S_{j,\bbeta}$ in a M-DCSBM is 
\begin{equation}
	\PP\left[A_{i,\balpha}^{j,\bbeta}=1\right] = \sigma_{i,\balpha}^{\bbeta}\, W^{s,\bbeta}_{r,\balpha}\, \sigma_{\balpha}^{j,\bbeta} \, ,
\label{eq:M-DCSBM-edge}
\end{equation}
where $W^{s,\bbeta}_{r,\balpha}$ is the expected number of edges from state nodes in layer $\balpha$ and \mesostructure{} $\set{S}_r$ to state nodes in layer $\bbeta$ and \mesostructure{} $\set{S}_s$, the quantity $\sigma^{\bbeta}_{i,\balpha}$ is the probability for an edge starting in \mesostructure{} $\set{S}_r$ in layer $\balpha$ and ending in layer $\bbeta$ to be attached to state node $(i,\balpha)$ (note that the dependence on $\set{S}_r$ is implicit in $\sigma^{\bbeta}_{i,\balpha}$), and $\sigma_{\balpha}^{j,\bbeta}$ is the probability for an edge starting in layer $\balpha$ and ending in \mesostructure{} $\set{S}_s$ in layer $\bbeta$ to be attached to state node $(j,\bbeta)$. (Note that the dependence on $\set{S}_s$ is implicit in $\sigma_{\balpha}^{j, \bbeta}$.) For an undirected M-DCSBM, both the block tensor $\mat{W}$ and the state-node parameters $\mat{\sigma}$ are symmetric. That is, $W^{r,\balpha}_{s,\bbeta}=W_{r,\balpha}^{s,\bbeta}$ and $\sigma^{i,\balpha}_{\bbeta} = \sigma_{i,\balpha}^{\bbeta}$.

The above M-DCSBM can generate multilayer networks with arbitrary expected layer-specific in-degrees and out-degrees for each state node. (The DCSBM of~\cite{Karrer2011} can generate single-layer networks with arbitrary expected degrees.) Given a multilayer network with adjacency tensor $\mat{A}$, the \define{layer-$\balpha$-specific in-degree} of state node $(j,\bbeta)$ is 
\begin{equation*}
	k^{j,\bbeta}_{\balpha} = \sum_{i \in \set{V}} A_{i,\balpha}^{j,\bbeta}\, ,
\end{equation*} 
and the \define{layer-$\bbeta$-specific out-degree} of state node $(i,\balpha)$ is
\begin{equation*}
	k_{i,\balpha}^{\bbeta} = \sum_{j \in \set{V}} A_{i,\balpha}^{j,\bbeta}\,.
\end{equation*}
The layer-$\balpha$-specific in-degree and out-degree of state node $(i,\balpha)$ are the \quoting{intralayer in-degree} and \quoting{intralayer out-degree} of $(i,\balpha)$.  For an undirected multilayer network, layer-$\bbeta$-specific in-degrees and out-degrees are equal (i.e., $k^{i,\balpha}_{\bbeta}=k_{i,\balpha}^{\bbeta}$). We refer to their common value as the ``layer-$\bbeta$-specific degree'' of a state node. For an ensemble of networks generated from an M-DCSBM, the associated means are
\begin{equation}
	\mean{k^{j,\bbeta}_{\balpha}}=\sigma^{j,\bbeta}_{\balpha} \sum_{r=1}^{|\set{S}|} W_{r,\balpha}^{s,\bbeta}\,,\quad s= S_{j,\bbeta}
\end{equation}
and
\begin{equation}
	\mean{k_{i,\balpha}^{\bbeta}}=\sigma_{i,\balpha}^{\bbeta} \sum_{s=1}^{|\set{S}|} W_{r,\balpha}^{s,\bbeta}\,,\quad r= S_{i,\balpha}\, .
\end{equation}

As we mentioned in \cref{sec:sampling_networks}, we consider undirected multilayer networks with only intralayer edges for our experiments in \cref{sec:numerical_examples}. The block tensor $\mat{W}$ thus does not have any interlayer contributions (i.e., $W_{r, \balpha}^{s,\bbeta}=0$ if $\balpha \neq \bbeta$). Furthermore, we can reduce the number of node parameters that we need to specify the M-DCSBM to a single parameter $\sigma_{i,\balpha}=\sigma_{i,\balpha}^{\balpha}=\sigma_{\balpha}^{i,\balpha}$ for each state node $(i,\balpha)$.

We sample the expected intralayer degrees ${e_{i,\balpha}=\langle k_{i,\balpha}^{\balpha}\rangle}$, where ${k_{i,\balpha}^{\balpha} = \sum_{j \in \set{V}_M} A_{i,\balpha}^{j,\balpha}}$, for the state nodes from a truncated power law 
\footnote{A power law with cutoffs $x_{\min}$ and $x_{\max}$ and exponent $\tau$ is a continuous probability distribution with probability density function 
\begin{equation*}
	p(x) = \begin{cases}  
		C\,x^{-\eta}\,, &x_{\min}\leq x \leq x_{\max}\,,\\
		0\,, & \text{otherwise\,,}
	\end{cases} 
\end{equation*}
where 
\begin{equation*}
	C=\frac{\eta-1}{x_{\min}^{-(\eta-1)} - x_{\max}^{-(\eta-1)}}
\end{equation*}
is the normalization constant.}
with exponent $\eta_k$, minimum cutoff $k_{\min}$, and maximum cutoff $k_{\max}$. 
We then construct the block tensor $\mat{W}$ and state-node parameters $\vec{\sigma}$ for the M-DCSBM from the sampled expected degrees $\vec{e}$ and the \mesostructure{} assignments $\set{S}$. Let 
\begin{equation*}
	\kappa_{s,\balpha} = \sum_{i \in \set{S}_s|_\balpha} e_{i,\balpha}\,, \qquad \set{S}_s \in \set{S}
\end{equation*} 
be the expected degree of \mesostructure{} $s$ in layer $\balpha$; and let 
\begin{equation*}
	w_{\balpha} =\frac{1}{2}\sum_{i\in \set{V}} e_{i,\balpha}
\end{equation*} 
be the expected number of edges in layer $\balpha$. Consequently,
\begin{equation*}
	\sigma_{i,\balpha} = \frac{e_{i,\balpha}}{\kappa_{s,\balpha}}\,, \qquad s=S_{i,\balpha}
\end{equation*}
is the probability for an intralayer edge in layer $\balpha$ to be attached to 
state node $(i, \balpha)$, given that the edge is attached to a state node that is in layer $\balpha$ and is part of the
community $S_{i,\balpha}$. 

The elements 
\begin{equation}\label{block-tensor}
	W_{r,\balpha}^{s,\bbeta}= \delta(\balpha,\bbeta)\left( (1-\mu) \delta(r,s) \kappa_{s,\balpha} + \mu \frac{\kappa_{r,\balpha} \kappa_{s, \balpha}}{2w_{\balpha}}\right)
\end{equation}
of the block tensor give, for $r \neq s$, the expected number of edges between state nodes in \mesostructure{} $s$ in layer $\bbeta$ and state nodes in \mesostructure{} $r$ in layer $\balpha$. For $s=r$ and $\bbeta=\balpha$, the block-tensor element $W_{r,\balpha}^{s,\bbeta}$ 
gives twice the expected number of edges. One way to think of the DCSBM benchmark is that we categorize each edge that we want to sample as an intra-\mesostructure{} edge with probability $1-\mu$ or as a ``random edge'' (i.e., an edge that can be either an intra-\mesostructure{} edge or an inter-\mesostructure{} edge) with probability $\mu$. To sample an edge, we sample two state nodes (which we then join by an edge). We call these two state nodes the ``end points'' of the edge. We sample the two end points of an intra-\mesostructure{} edge with a frequency that is proportional to the expected degree of their associated state nodes, conditional on the end points being in the same \mesostructure{}. By contrast, we sample the two end points of a random edge with a frequency proportional to the expected degree of their associated state nodes (without conditioning on \mesostructure{} assignment). We assume that the total number of edges in layer $\balpha$ is sampled from a Poisson distribution~\footnote{The choice of a Poisson distribution for the number of edges ensures that this approach for sampling edges is approximately consistent with sampling edges independently from Bernoulli distributions with success probabilities 
\cref{eq:M-DCSBM-edge}. The exact distribution for the number of edges is a Poisson--binomial distribution, from which it is difficult to sample; a Poisson--binomial distribution is well-approximated by a Poisson distribution if none of the individual edge probabilities are too large.} with mean $w_{\balpha}$. Although our procedure for sampling edges describes a potential algorithm for sampling networks from the DCSBM benchmark, it usually is more efficient to sample edges separately for each pair of \mesostructure{}s. 

In \cref{alg:dcsbm}, we show pseudocode for the specific instance of the model in \cref{eq:M-DCSBM-edge} that we use to sample intralayer network edges (independently for each induced partition) for a given multilayer partition in our numerical experiments of \cref{sec:numerical_examples}. The only difference between \cref{alg:dcsbm} and the sampling algorithm of~\cite{Karrer2011} is that we use rejection sampling to avoid creating self-edges and multi-edges. (In other words, if we sample an edge that has already been sampled or that is a self-edge, we do not include it in the multilayer network; instead, we resample.) Rejection sampling is efficient provided all blocks of the network are sufficiently sparse, such that the probability of generating multi-edges is small. For dense blocks of a network, we instead sample edges from independent Bernoulli distributions with success probability \cref{eq:M-DCSBM-edge}. This algorithm for sampling networks from a DCSBM is very efficient; it scales linearly with the number of edges in a network. 

In the above discussion, we defined the parameters in  \cref{eq:M-DCSBM-edge} to generate intralayer edges in a multilayer network by specifying layer-specific in-degrees and out-degrees. That is, the layer-$\bbeta$-specific in-degree and out-degree of state node $(i,\balpha)$ are $0$ if $\bbeta\neq\balpha$. We can also extend the model in \cref{eq:M-DCSBM-edge} to generate interlayer edges. For a directed multilayer network with interlayer edges, we sample expected layer-$\bbeta$-specific in-degrees $e_{\bbeta}^{i, \balpha}$ and out-degrees $e_{i,\balpha}^{\bbeta}$ for each state node $(i, \balpha)$ and layer $\bbeta$ from appropriate distributions. Given expected layer-specific degrees $\mat{e}$ and a multilayer partition $\set{S}$, we then construct the block tensor $\mat{W}$ and state-node parameters $\mat{\sigma}$ for the M-DCSBM analogously to the special case that we described above. Let 
\begin{equation*}
	\kappa^{s, \balpha}_{\bbeta}= \sum_{i \in \set{S}_s|_{\balpha}} e^{i,\balpha}_{\bbeta}\,,\quad
\kappa_{s, \balpha}^{\bbeta}= \sum_{i \in \set{S}_s|_{\balpha}} e_{i,\balpha}^{\bbeta}\,,\quad
\set{S}_s \in \set{S}
\end{equation*}
be the expected layer-$\bbeta$-specific in-degree and out-degree of \mesostructure{} $s$ in layer $\balpha$; and let 
\begin{equation*}
	w_{\balpha}^{\bbeta} = \sum_{i \in \set{V}} e_{i, \balpha}^{\bbeta}=\sum_{i \in \set{V}}  e_{\balpha}^{i,\bbeta}
\end{equation*}
be the expected number of edges from layer $\balpha$ to layer $\bbeta$. (Note the consistency constraint on expected in-degrees and expected out-degrees.) It then follows that 
\begin{equation*}
	\sigma_{\bbeta}^{i,\balpha} = \frac{e_{\bbeta}^{i,\balpha}}{\kappa_{\bbeta}^{s,\balpha}}\,,\quad
\sigma^{\bbeta}_{i,\balpha} = \frac{e^{\bbeta}_{i,\balpha}}{\kappa^{\bbeta}_{s,\balpha}}\,,\quad 
s =S_{i,\balpha}\,,
\end{equation*}
and 
\begin{equation*}
	W_{r, \balpha}^{s,\bbeta} = (1-\mu)\delta(r,s) \frac{\kappa_{s,\balpha}^{\bbeta} + \kappa_{\balpha}^{s,\bbeta}}{2} + \mu \frac{\kappa_{r,\balpha}^{\bbeta} \kappa_{\balpha}^{s,\bbeta}}{w_{\balpha}^{\bbeta}}\,.
\end{equation*} For directed networks, the expected layer-specific in-degrees and out-degrees generated by the above model do not correspond exactly to the values 
from the input parameters $\mat{e}$, except when $\mu=1$.

%%%%%%%%%%%%%%%%%%%%%%%%%%%%%%%%
%%%%%%%%%%%%%%%%%%%%%%%%%%%%%%%%

\section{The Louvain algorithm and its variants}
\label{app:louvain}

In this appendix, we describe the Louvain algorithm and variants of the Louvain algorithm that we use in the numerical experiments in \cref{sec:numerical_examples}. 

The Louvain algorithm \cite{Blondel2008} for maximizing (single-layer or multilayer) modularity proceeds in two phases, which are repeated iteratively. Starting from an initial partition, one considers the state nodes one by one (in some order) and places each state node in a set that results in the largest increase of modularity. (If there is no move that improves modularity, a state node keeps the same assignment.) One repeats this first phase of the algorithm until reaching a local maximum. In the second phase of the Louvain algorithm, one obtains a new 
modularity matrix by aggregating the sets of state nodes that one obtains after the convergence of the first phase. 
One then applies the algorithm's first phase to the new modularity matrix and iterates both phases until one converges to a local maximum. The two Louvain-like algorithms that we use in \cref{sec:numerical_examples} differ in how they select which moves to make. The first is \textsc{GenLouvain}, which always chooses the move that maximally increases modularity; the second is \textsc{GenLouvainRand}, which chooses modularity-increasing moves at random, such that the probability of a particular move is proportional to the resulting increase in the quality function. (The latter is a variant of the algorithm \quoting{LouvainRand} in \cite{Bazzi2014}; in that algorithm, one chooses modularity-increasing moves uniformly at random.) We use \todefine{reiteration} and \todefine{post-processing} to improve the output of \textsc{GenLouvainRand}. Reiteration entails taking an output partition of \textsc{GenLouvainRand} as an initial partition and reiterating the algorithm until the output partition no longer changes. Post-processing entails using the Hungarian algorithm~\cite{Kuhn1955} to optimize the \quoting{persistence}~\cite{Bazzi2014} (i.e., the number of interlayer edges for which both end points have the same community assignment) of each output partition without changing the induced partitions on each layer. All Louvain variants and related functionality are available at~\cite{genlouvain}. 

%%%%%%%%%%%%%%%%%%%%%%%%%%%%%%%%
%%%%%%%%%%%%%%%%%%%%%%%%%%%%%%%%

\section{Multilayer NMI experiments}
\label{app:mNMI}

\begin{figure}[t!]
\centering
\includegraphics{fig/temporalCP_colorbar_L.pdf}\\
\begin{tabular}{@{}p{\linewidth}@{}}
\textsc{GenLouvain}\\
\hline
\vspace{-1.5em}
\subfloat[$\hat{p}=0.85,$]{
\includegraphics{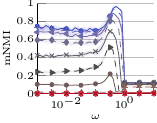}%
}%
\hfill
\subfloat[$\hat{p}=0.95$]{
\includegraphics{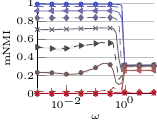}%
}%
\hfill
\subfloat[$\hat{p}=1$]{
\includegraphics{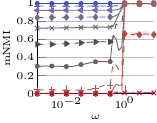}%
}\\
\vspace{0em}
\textsc{GenLouvainRand, no PP}\\
\hline
\vspace{-1.5em}
\subfloat[$\hat{p}=0.85$]{
\includegraphics{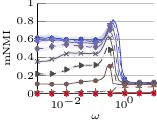}%
}%
\hfill
\subfloat[$\hat{p}=0.95$]{
\includegraphics{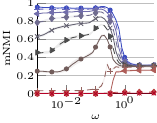}%
}%
\hfill
\subfloat[$\hat{p}=1$]{
\includegraphics{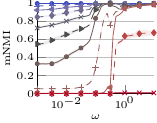}%
}\\
\vspace{0em}
\textsc{GenLouvainRand}\\
\hline
\vspace{-1.5em}
\subfloat[$\hat{p}=0.85$]{
\includegraphics{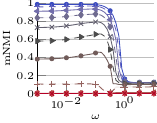}%
}%
\hfill
\subfloat[$\hat{p}=0.95$]{
\includegraphics{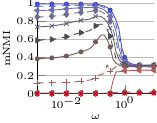}%
}%
\hfill
\subfloat[$\hat{p}=1$]{
\includegraphics{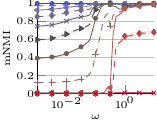}%
}\\
\vspace{0em}
\textsc{InfoMap}\\
\hline
\vspace{-1.5em}
\subfloat[$\hat{p}=0.85$]{
\includegraphics{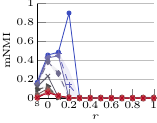}%
}%
\hfill
\subfloat[$\hat{p}=0.95$]{
\includegraphics{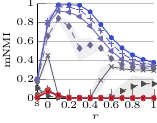}%
}%
\hfill
\subfloat[$\hat{p}=1$]{
\includegraphics{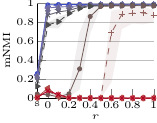}%
}%
\end{tabular}
\caption{Effect of interlayer coupling strength $\omega$ and relaxation rate $r$ on the ability of different community-detection algorithms to recover planted partitions as a function of the community-mixing parameter $\mu$ in a uniform multiplex benchmark (see \cref{sfig:multiplex}). Each multilayer network has $1000$ nodes and $15$ layers, and each node is present in all layers. 
\label{fig:mNMImult-bench}}
\end{figure}

\begin{figure}[t!]
\centering
\includegraphics{fig/temporalCP_colorbar_L.pdf}\\
\begin{tabular}{@{}p{\linewidth}@{}}
\textsc{GenLouvain}\\
\hline
\vspace{-1.5em}
\subfloat[$\hat{p}=0.85$, $p_c = 0$]{
\includegraphics{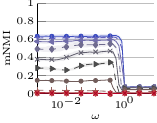}%
}%
\hfill
\subfloat[$\hat{p}=0.95$, $p_c = 0$]{
\includegraphics{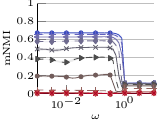}%
}%
\hfill
\subfloat[$\hat{p}=1$, $p_c = 0$]{
\includegraphics{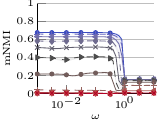}%
}\\
\vspace{0em}
\textsc{GenLouvainRand, no PP}\\
\hline
\vspace{-1.5em}
\subfloat[$\hat{p}=0.85$, $p_c = 0$]{
\includegraphics{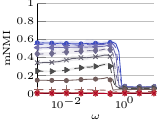}%
}%
\hfill
\subfloat[$\hat{p}=0.95$, $p_c = 0$]{
\includegraphics{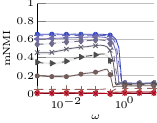}%
}%
\hfill
\subfloat[$\hat{p}=1$, $p_c = 0$]{
\includegraphics{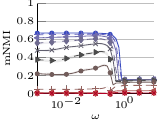}%
}\\
\vspace{0em}
\textsc{GenLouvainRand}\\
\hline
\vspace{-1.5em}
\subfloat[$\hat{p}=0.85$, $p_c = 0$]{
\includegraphics{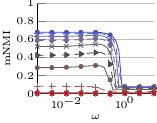}%
}%
\hfill
\subfloat[$\hat{p}=0.95$, $p_c = 0$]{
\includegraphics{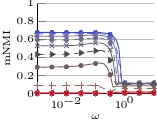}%
}%
\hfill
\subfloat[$\hat{p}=1$, $p_c = 0$]{
\includegraphics{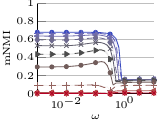}%
}\\
\vspace{0em}
\textsc{InfoMap}\\
\hline
\vspace{-1.5em}
\subfloat[$\hat{p}=0.85$, $p_c = 0$]{
\includegraphics{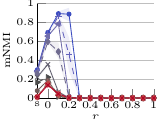}%
}%
\hfill
\subfloat[$\hat{p}=0.95$, $p_c = 0$]{
\includegraphics{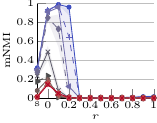}%
}%
\hfill
\subfloat[$\hat{p}=1$, $p_c  = 0$]{
\includegraphics{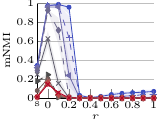}%
}%
\end{tabular}
\caption{Effect of interlayer coupling strength $\omega$ and relaxation rate $r$ on the ability of different community-detection algorithms to recover planted partitions as a function of the community-mixing parameter $\mu$ in a multiplex benchmark with nonuniform interlayer dependencies. Each multilayer network has $1000$ nodes and $15$ layers, and each node is present in all layers. The layer-coupled interlayer-dependency matrix is a block-diagonal matrix with diagonal blocks of size $5\times 5$. For each diagonal block, we set the value in the interlayer-dependency matrix to a value $p$ (so each diagonal block has the same structure as the matrix in \cref{sfig:multiplex}); for each off-diagonal block, we set the value to $p_c = 0$, thereby incorporating an abrupt change in community structure. We define the null distributions in these experiments so that there is no overlap in community labels between different groups of layers.
\label{fig:mNMImulthet-bench}}
\end{figure}

\begin{figure}[t!]
\centering
\includegraphics{fig/temporalCP_colorbar_L.pdf}\\
\begin{tabular}{@{}p{\linewidth}@{}}
\textsc{GenLouvain}\\
\hline
\vspace{-1.5em}
\subfloat[$p=0.85$]{
\includegraphics{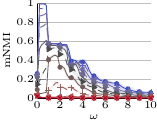}%
}%
\hfill
\subfloat[$p=0.95$]{
\includegraphics{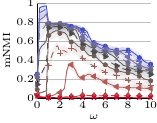}%
}%
\hfill
\subfloat[$p=1$]{
\includegraphics{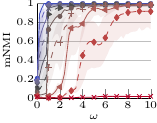}%
}\\
\vspace{0em}
\textsc{GenLouvainRand, no PP}\\
\hline
\vspace{-1.5em}
\subfloat[$p=0.85$]{
\includegraphics{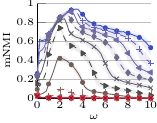}%
}%
\hfill
\subfloat[$p=0.95$]{
\includegraphics{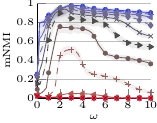}%
}%
\hfill
\subfloat[$p=1$]{
\includegraphics{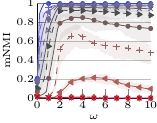}%
}\\
\vspace{0em}
\textsc{GenLouvainRand}\\
\hline
\vspace{-1.5em}
\subfloat[$p=0.85$]{
\includegraphics{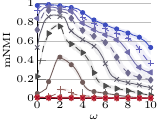}%
}%
\hfill
\subfloat[$p=0.95$]{
\includegraphics{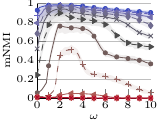}%
}%
\hfill
\subfloat[$p=1$]{
\includegraphics{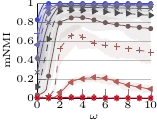}%
}\\
\vspace{0em}
\textsc{InfoMap}\\
\hline
\vspace{-1.5em}
\subfloat[$p=0.85$]{
\includegraphics{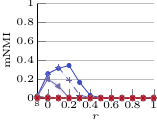}%
}%
\hfill
\subfloat[$p=0.95$]{
\includegraphics{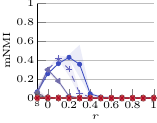}%
}%
\hfill
\subfloat[$p=1$]{
\includegraphics{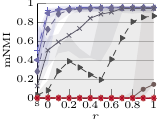}%
}%
\end{tabular}
\caption{Effect of interlayer coupling strength $\omega$ and relaxation rate $r$ on the ability of different community-detection algorithms to recover planted partitions as a function of the community-mixing parameter $\mu$ in a temporal benchmark with uniform interlayer dependencies (i.e., $p_\beta = p \in[0,1]$ for all $\beta\in\{2,\ldots,l\}$ in \cref{sfig:temporal}). Each multilayer network has $150$ nodes and $100$ layers, and each node is present in all layers.
\label{fig:mNMItemp-bench}}
\end{figure}

\begin{figure}[t!]
\centering
\includegraphics{fig/temporalCP_colorbar_L.pdf}\\
\begin{tabular}{@{}p{\linewidth}@{}}
\textsc{GenLouvain}\\
\hline
\vspace{-1.5em}
\subfloat[$p=0.85$, $p_c = 0$]{
\includegraphics{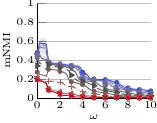}%
}%
\hfill
\subfloat[$p=0.95$, $p_c = 0$]{
\includegraphics{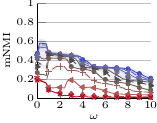}%
}%
\hfill
\subfloat[$p=1$, $p_c = 0$]{
\includegraphics{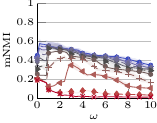}%
}\\
\vspace{0em}
\textsc{GenLouvainRand, no PP}\\
\hline
\vspace{-1.5em}
\subfloat[$p=0.85$, $p_c = 0$]{
\includegraphics{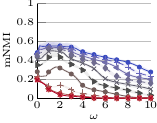}%
}%
\hfill
\subfloat[$p=0.95$, $p_c = 0$]{
\includegraphics{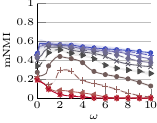}%
}%
\hfill
\subfloat[$p=1$, $p_c = 0$]{
\includegraphics{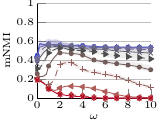}%
}\\
\vspace{0em}
\textsc{GenLouvainRand}\\
\hline
\vspace{-1.5em}
\subfloat[$p=0.85$, $p_c = 0$]{
\includegraphics{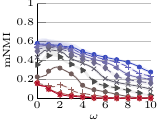}%
}%
\hfill
\subfloat[$p=0.95$, $p_c = 0$]{
\includegraphics{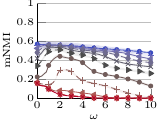}%
}%
\hfill
\subfloat[$p=1$, $p_c = 0$]{
\includegraphics{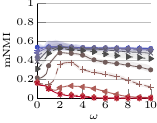}%
}\\
\vspace{0em}
\textsc{InfoMap}\\
\hline
\vspace{-1.5em}
\subfloat[$p=0.85$, $p_c = 0$]{
\includegraphics{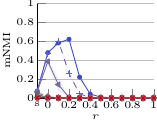}%
}%
\hfill
\subfloat[$p=0.95$, $p_c = 0$]{
\includegraphics{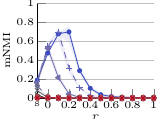}%
}%
\hfill
\subfloat[$p=1$, $p_c = 0$]{
\includegraphics{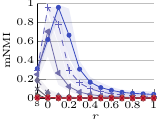}%
}%
\end{tabular}
\caption{Effect of interlayer coupling strength $\omega$ and relaxation rate $r$ on the ability of different community-detection algorithms to recover planted partitions as a function of the community-mixing parameter $\mu$ in a temporal benchmark with nonuniform interlayer dependencies. Each multilayer network has $150$ nodes and $100$ layers, and each node is present in all layers. Every 25th layer (i.e., for layers 25, 50, and 75), we set the value of $p_\beta$ in \cref{sfig:temporal} to $p_c = 0$ (thereby incorporating an abrupt change in community structure); we set all other values of $p_\beta$ in \cref{sfig:temporal} to $p$. We define the null distributions in these experiments so that there is no overlap in community labels between different groups of layers.  
\label{fig:mNMItempCP-bench}}
\end{figure}

In this appendix, we calculate the multilayer NMI (mNMI) between the multilayer output partition and the planted multilayer partition for the numerical experiments in \cref{fig:mult-bench,fig:Mhet-bench,fig:temp-bench,fig:tempCP-bench}. One needs to be cautious when interpreting values of mNMI; they depend on how one determines the correspondences between communities in different layers, and these correspondences can be ambiguous. For example, in a temporal network, if a community in one layer splits into multiple communities in the next layer, there are multiple plausible ways to define labels across layers (e.g., all new communities get new labels or the largest new community keeps the original label). Moreover, different quality functions reward interlayer-assignment labeling in different ways, and mNMI conflates this issues with how well a quality function recovers structure within layers. In particular, a small value of mNMI can hide an optimal value of $\langle \text{NMI} \rangle$ (as $\langle \text{NMI} \rangle=1$ is necessary, but not sufficient, to have $\text{mNMI}=1$).

We include only a subset of the values of $p$ and $\hat{p}$ that we considered in \cref{fig:mult-bench,fig:temp-bench}, as these suffice for the purpose of this appendix. In the numerical experiments of \cref{sec:numerical_examples}, we observed that post-processing does not have a major effect on $\mean{\NMI}$. This suggests that any misassignment across layers that underemphasizes the \quoting{persistence} of a multilayer partition (which post-processing tries to mitigate~\cite{Bazzi2014}) does not affect an algorithm's ability to identify structure within layers. (Recall that post-processing relabels an assignment without changing the induced partitions.) We include figures without post-processing in the experiments of this appendix, as we expect the effect to be more noticeable when examining NMI values between multilayer partitions (which accounts for assignments both within and across layers).  

In \cref{fig:mNMImult-bench,fig:mNMImulthet-bench}, we use the same benchmark networks as in the multiplex examples
in \cref{fig:mult-bench,fig:Mhet-bench}, respectively. In \cref{fig:mNMItemp-bench,fig:mNMItempCP-bench}, we use the same benchmark networks as in the temporal examples of \cref{fig:temp-bench,fig:tempCP-bench}, respectively. All mNMI values are means over $10$ runs of the algorithms and $100$ instantiations of the benchmark. (See the introduction of \cref{sec:numerical_examples} for more details on how we generate these instantiations.) Each curve in each figure corresponds to the mean mNMI values that we obtain for a given value of the community-mixing parameter $\mu$, and the shaded area around a curve corresponds to the minimum and maximum mNMI values that we obtain with the $10$ sample partitions for a given value of $p$ or $\hat{p}$. 

For experiments with nonuniform interlayer dependencies in \cref{fig:mNMItempCP-bench,fig:mNMImulthet-bench}, it seems that \textsc{InfoMap} is better at detecting abrupt differences (``breaks'', such as in the form of change points for temporal examples) in community structure
than both \textsc{GenLouvain} and \textsc{GenLouvainRand}, especially for the multiplex case (where this manifests for all examined values of $\hat{p}$). However, multilayer \textsc{InfoMap} correctly identifies the planted community structure only when it is particularly strong (specifically, for $\mu\leq 0.3$).
Recall from our experiments in \cref{fig:Mhet-bench,fig:tempCP-bench} that both \textsc{GenLouvain} and \textsc{GenLouvainRand} are better than \textsc{InfoMap} at detecting an induced partition within a layer (especially in cases where the planted community structure is quite weak). The reason 
that multilayer modularity maximization does not perfectly recover breaks in community structure of the planted multilayer partition
is because multilayer modularity maximization with uniform ordinal or categorical coupling (see the introduction of \cref{sec:numerical_examples}) incentivizes ``persistence'' of community labels between layers (and nonzero persistence is achievable even when one independently samples induced partitions on different layers)~\cite{Bazzi2014}. However, by design, our benchmark networks have no persistence whenever there is a break in community structure because of how we defined the support of the null distributions. Therefore, we do not expect multilayer modularity maximization to recover the planted labels between layers when $p_c=0$; this manifests as a de facto maximum in the mNMI between the output partition and the planted partition.

%%%%%%%

\section*{Acknowledgements}

MB acknowledges a CASE studentship award from the EPSRC (BK/10/41). MB was also supported by The Alan Turing Institute under the EPSRC grant EP/N510129/1. LGSJ acknowledges a CASE studentship award from the EPSRC (BK/10/39). MB, LGSJ, AA, and MAP were supported by FET-Proactive project PLEXMATH (FP7-ICT-2011-8; grant \#317614) funded by the European Commission; and MB, LGSJ, and MAP were also supported by the James S. McDonnell Foundation (\#220020177). MAP was also supported by the National Science Foundation (grant \#1922952) through the Algorithms for Threat Detection (ATD) program. Computational resources that we used for this research were supported in part by the National Science Foundation under Grant No. CNS-0521433, in part by Lilly Endowment, Inc. through its support for the Indiana University Pervasive Technology Institute, and in part by the Indiana METACyt Initiative. The Indiana METACyt Initiative at Indiana University Bloomington was also supported in part by Lilly Endowment, Inc. We thank Sang Hoon Lee, Peter Mucha (and his research group), Brooks Paige, and Roxana Pamfil for helpful comments.

%%%%%%%

%\bibliography{Bibliography2}

\begin{thebibliography}{134}
\bibitem [{not({\natexlab{a}})}]{note_lexi}%
  \BibitemOpen
  \bibinfo {note} {One defines lexicographic order for the Cartesian product of
  ordered sets similarly to the way that words are ordered in a dictionary. For
  two elements $e,f \in \set{E}=\set{E}_1\times\ldots\times \set{E}_n$, we
  define $e\prec f$ if ${e_1<f_1}$ or there exists $i$ such that $e_{i} <
  f_{i}$ and $e_{i'}=f_{i'}$ for all $i' < i$.}\BibitemShut {Stop}%
\end{thebibliography}


%%%%%%

\end{document}